%% file: 2022-sigir-fp-convqa.tex
\documentclass[sigconf,natbib=true,anonymous=false]{acmart}

\AtBeginDocument{%
  \providecommand\BibTeX{{%
    \normalfont B\kern-0.5em{\scshape i\kern-0.25em b}\kern-0.8em\TeX}}}

\usepackage{latexsym}
\usepackage{graphicx}
\graphicspath{{./images/}}
\usepackage{booktabs} 
\usepackage{color}  
\usepackage{amsmath}  
\usepackage{subcaption}
\usepackage{caption}
\usepackage{tikz}
\usepackage{colortbl} 
\usepackage{framed}
\usepackage{multirow}
\usepackage{multicol}
\usepackage{hyperref}
\usepackage{url}
\usepackage{balance}
\usepackage{verbatim}
\usepackage{cancel}
\usepackage{xspace} 
\usepackage[ruled,vlined]{algorithm2e}
\usepackage{bbold} 
\usepackage{balance}

\newcommand{\struct}[1]{\texttt{\small #1}}
\newcommand{\utterance}[1]{\textit{#1}}
\newcommand{\phrase}[1]{\textit{``#1''}}

\newenvironment{Snugshade}[1][236,236,236]{
    \setlength{\itemsep}{0pt}
     \setlength{\parsep}{0pt}
     \setlength{\topsep}{0pt}
     \setlength{\partopsep}{0pt}
     \setlength{\leftmargin}{1.5em}
     \setlength{\labelwidth}{0em}
     \setlength{\labelsep}{0em} 
    \setlength{\parskip}{0pt}
    \definecolor{shadecolor}{RGB}{#1}%
    \begin{snugshade}
}{%
    \end{snugshade}%
}

\newcommand{\convinse}{\textsc{Convinse}\xspace}
\newcommand{\convmix}{\textsc{ConvMix}\xspace}

\newcommand{\clocq}{\textsc{Clocq}\xspace}

\newcommand{\quretec}{\textsc{QuReTeC}\xspace}

\acmConference[SIGIR '22]{Proceedings of the 45th International ACM SIGIR Conference on Research and Development in Information Retrieval}{July 11--15, 2022}{Madrid, Spain}

\begin{document}


\title{Conversational Question Answering on Heterogeneous Sources}

\author{Philipp Christmann}
\affiliation{%
  \institution{Max Planck Institute for Informatics}
  \streetaddress{Saarland Informatics Campus}
  \city{Saarbruecken}
 \country{Germany}
}
\email{pchristm@mpi-inf.mpg.de}

\author{Rishiraj Saha Roy}
\affiliation{%
  \institution{Max Planck Institute for Informatics}
  \streetaddress{Saarland Informatics Campus}
  \city{Saarbruecken}
    \country{Germany}
}
\email{rishiraj@mpi-inf.mpg.de}

\author{Gerhard Weikum}
\affiliation{%
  \institution{Max Planck Institute for Informatics}
  \streetaddress{Saarland Informatics Campus}
  \city{Saarbruecken}
    \country{Germany}
}
\email{weikum@mpi-inf.mpg.de}

\renewcommand{\shortauthors}{P. Christmann et al.}

\newcommand{\squishlist}{
    \begin{list}{$\bullet$}{ 
        \setlength{\itemsep}{0pt}
        \setlength{\parsep}{1pt}
        \setlength{\topsep}{1pt}
        \setlength{\partopsep}{0pt}
        \setlength{\leftmargin}{1.5em}
        \setlength{\labelwidth}{1em}
        \setlength{\labelsep}{0.5em} 
    } 
}

\newcommand{\squishend}{
  \end{list}  }
  
\newcommand{\GW}[1]{{\color{blue}{GW: #1}} }
\newcommand{\PC}[1]{{\color{orange}{PC: #1}} }

\newcommand{\myparagraph}[1]{\noindent \textbf{#1}.}

\setcounter{secnumdepth}{4}

\input{sections/00-abstract}



\maketitle
\input{sections/01-introduction}
\input{sections/02-concepts}
\input{sections/03-method}
\input{sections/04-benchmark}

\input{sections/05-experiments}
\input{sections/06-results}

\input{sections/07-related}

\input{sections/08-conclusion}


\balance

\bibliographystyle{ACM-Reference-Format}
\bibliography{convinse}

\end{document}

%% file: sections/00-abstract.tex
\begin{abstract}


Conversational question answering (ConvQA) tackles sequential information needs where contexts in follow-up questions are left implicit. Current ConvQA systems operate over homogeneous sources of information: either a knowledge base (KB), or a text corpus, or a collection of tables. This paper addresses the novel issue of jointly tapping into all of these together, this way boosting answer coverage
and confidence. We present \convinse, an end-to-end pipeline for ConvQA over heterogeneous sources, operating in three stages: i) learning an explicit structured representation of an incoming question and its conversational context, ii) harnessing this frame-like representation to uniformly capture relevant evidences from KB, text, and tables, and iii) running a fusion-in-decoder model to generate the answer. We construct and release the first benchmark, \convmix, for ConvQA over heterogeneous sources, comprising $3000$ real-user conversations with $16000$ questions, along with entity annotations, completed question utterances, and question paraphrases. Experiments demonstrate the viability and advantages of our method, compared to state-of-the-art baselines.
\end{abstract}

%% file: sections/01-introduction.tex
\section{Introduction}
\label{sec:intro}


\myparagraph{Motivation} Conversational question answering (ConvQA)~\cite{choi2018quac,saharoy2021question,reddy2019coqa,qu2020open} is
is a popular mode of communication with digital personal assistants like Alexa, Cortana, Siri, or the Google Assistant, that are ubiquitous in today's devices. In ConvQA, users pose questions to the system sequentially, over multiple turns. In conversations between two humans, follow-up questions
usually contain \textit{implicit} context. 
The ConvQA system is expected to resolve such implicit information from the conversational history.
Consider, for example, a typical ConvQA session on factual knowledge below:
\begin{Snugshade}
    \utterance{$q^0$: Who played Jaime Lannister in GoT?}\\
    \indent $a^0$: \struct{Nikolaj Coster-Waldau}\\
    \indent \utterance{$q^1$: What about the dwarf?}\\
    \indent $a^1$: \struct{Peter Dinklage}\\
    \indent \utterance{$q^2$: When was he born?}\\
    \indent $a^2$: \struct{11 June 1969}\\
    \indent \utterance{$q^3$: Release date of first season?}\\
    \indent $a^3$: \struct{17 April 2011}\\
    \indent \utterance{$q^4$: Duration of an episode?}\\
    \indent $a^4$: \struct{50-82 minutes}
\end{Snugshade}
State-of-the-art works on ConvQA make use of single information sources: either curated knowledge bases (KB)~\cite{christmann2019look,guo2018dialog,kaiser2021reinforcement,shen2019multi,saha2018complex,kacupaj2021conversational,plepi2021context,marion2021structured,lan2021modeling}, or unstructured text collections~\cite{chengraphflow,huang2018flowqa,qiu2021reinforced,qu2019attentive,qu2019bert}, or 
Web
tables~\cite{mueller2019answering,iyyer2017search},
but only one of these.
Questions $q^0$ and $q^2$ can be answered more conveniently using KBs like Wikidata~\cite{vrandevcic2014wikidata}, YAGO~\cite{suchanek2007yago}, or DBpedia~\cite{auer2007dbpedia}, that store factual world knowledge in compact RDF triples. However, answering $q^1$, $q^3$ or $q^4$
via a KB
requires complex reasoning efforts.
For $q^1$, even with named entity disambiguation (NED) in conversations~\cite{shang2021entity,joko2021conversational}, it is unlikely that the correct KB entity (\struct{Tyrion Lannister}) can be inferred, which means that the resulting answer search space~\cite{christmann2022beyond} will not have the answer.
For $q^3$, answering via a KB requires a two-step lookup involving the first season, and then the corresponding release date.
For $q^4$, a KB might have the details for each individual episode, but collecting this information and aggregating for the final answer can be quite cumbersome. 
In contrast, the answers to these three questions are much more easily spotted in content of text documents, or Web tables.
In addition, there are obviously many information needs where answers are present only in text form, as KBs and Web tables have inherently limited coverage. An example would be a question like:
\utterance{What did Brienne call Jaime?} (\phrase{Kingslayer}).
A smart ConvQA system should, therefore,
be able to tap into more than one kind of knowledge repository, to improve answer recall and to
boost answer confidence by
leveraging multiple kinds of evidence across sources.

\begin{table} [t] \small 
    \caption{Question understanding approaches for ConvQA.}
    \vspace*{-0.3cm}
    \resizebox*{\columnwidth}{!}{
	\begin{tabular}{l l}
	    \toprule
	    \textbf{Original} $q^3$         & \utterance{Release date of first season?}                             \\ \midrule
	    \textbf{Question resolution}    & \utterance{Release date of first season? in GoT}                      \\
	    \textbf{Question rewriting}     & \utterance{What was the release date of the first season of GoT?}     \\
	    \textbf{\convinse SR}           & $\langle$ \struct{\textcolor{gray}{GoT} | \textcolor{red}{first season} | \textcolor{blue}{release date} | \textcolor{cyan}{date}} $\rangle$  \\
	    \bottomrule
	\end{tabular}}
	\label{tab:qu}
	\vspace*{-0.5cm}
\end{table}

\myparagraph{Limitations of state-of-the-art}
Existing research on ConvQA has considered solely one kind of information source for deriving answers.
Further, when specializing on a given source, methods often adopt source-specific design choices that do not generalize well~\cite{huang2018flowqa,guo2018dialog,christmann2019look}.
For example, representations of the conversational context, like KB subgraphs or text passages, are often specifically modeled for the knowledge repository at hand,
making these heterogeneous sources apparently incompatible.
Methods for question rewriting~\cite{vakulenko2021question,elgohary2019can,raposo2022question} and
question resolution~\cite{kumar2017incomplete,voskarides2020query}
convert
short user utterances into full-fledged
questions where the intent is made completely \textit{explicit}.
However, this 
adds major complexity and may lose
valuable cues from the conversation flow.
Further, these methods face evidence retrieval problems arising from long and potentially verbose questions~\cite{gupta2015information}.

There has been substantial work on single-question QA over heterogeneous
sources~\cite{xiong2019improving,chen2020open,hannan2020manymodalqa,oguz2021unikqa,sun2018open,sun2019pullnet,talmor2021multimodalqa,zhu2021tat}, with complete questions as input. Among these, only O\u{g}uz et al.~\cite{oguz2021unikqa} and Ma et al.~\cite{ma2021open}
try to deal with
KBs, text, and tables.
Their approach 
is designed for simple questions, though,
and cannot easily be extended to the challenging ConvQA setting ~\cite{saharoy2021question}.
Finally, none of the prior works on ConvQA produce human-interpretable structures that could assist end users in case of erroneous system responses.

\myparagraph{Approach}
To overcome these limitations, we propose
\convinse
(\underline{CONV}QA with \underline{IN}termediate Representations on Heterogeneous \underline{S}ources for \underline{E}xplainability),
an end-to-end framework for 
conversational QA
on
a mixture of sources.
\convinse consists of three main stages: i) \textit{question understanding (QU)}, ii) \textit{evidence retrieval and scoring (ERS)}, and iii) \textit{heterogeneous answering (HA)}.

The first stage, QU, is our primary contribution in this work.
It addresses the challenges of incomplete user utterances introduced by the conversational setting.
We derive an \textit{intent-explicit structured representation (SR)} that captures the complete information need. Table~\ref{tab:qu} shows such an SR for $q^3$ of our running example.
SRs are frame-like structures for a question that contain designated slots for open-vocabulary lexical representations of entities in the conversational context (marked \textcolor{gray}{gray}) and the current question (\textcolor{red}{red}), relational predicates (\textcolor{blue}{blue}), and expected answer types (\textcolor{cyan}{cyan}). 
SRs can be viewed as concise
gists of user intents, 
intended to be in a form
independent of any specific answering source.
They are
self-contained
interpretable 
representations of the 
user's information need,
and are inferred using 
fine-tuned transformer models trained on data generated by distant supervision from 
plain sequences of QA pairs.
We further propose a conversational flow graph (CFG), which can be inferred 
from the SR, and enhances the explainability of the derivation process.

The second stage, ERS,
exploits recent developments in entity-based retrieval~\cite{christmann2022beyond}
to judiciously retrieve question-relevant evidences (KB-facts, text-sentences, table-records, or infobox-entries) from each information source. These heterogeneous evidences are
verbalized~\cite{oguz2021unikqa,jia21complex,pramanik2021uniqorn} on-the-fly and run through a scoring model.
The top-$k$ pieces of evidence are passed to the answering stage.

The third and final stage, HA, consists of a fusion-in-decoder (FiD) model~\cite{izacard2021leveraging}, that is state-of-the-art in the retrieve-and-read paradigm for open-domain QA.
FiD acts as a ``generative reader'', creating a crisp answer from the top-$k$ evidences, that is returned to the end user.

\myparagraph{Benchmark} 
Another novel contribution is the construction of \convmix, the first benchmark for ConvQA over heterogeneous sources.
\convmix is a crowdsourced dataset that 
contains questions with answers emanating from the Wikidata KB, the full text of Wikipedia articles, and the collection of Wikipedia tables and infoboxes.
\convmix contains $2800$ conversations with five turns ($14$k utterances), and $200$ conversations with ten turns ($2$k utterances),
their gold answers and respective knowledge sources for answering. 
Conversations are accompanied by metadata like entity annotations, completed questions, and paraphrases. 
The collected dataset \convmix, and all our code and data for \convinse can be accessed at \textbf{\url{https://convinse.mpi-inf.mpg.de}}.

\vspace*{0.2cm}
\myparagraph{Contributions} Our salient contributions are the following:
\squishlist
    \item The paper proposes \convinse, the first end-to-end method for ConvQA over heterogeneous sources.
    \item It introduces structured representations to capture user intents in a structured and explainable manner, a key element for seamless answering over a mixture of heterogeneous sources.
    \item It presents distant supervision mechanisms to automatically annotate conversations
    with
    structured representations. 
    \item It provides \convmix, the first benchmark for ConvQA over heterogeneous sources.
\squishend


%% file: sections/02-concepts.tex
\section{Concepts and notation}
\label{sec:concepts}


\myparagraph{Question} A natural language question $q$ is a sequence of words expressing an interrogative intent. A question can be complete (i.e., self-contained/full-fledged/intent-explicit), or incomplete (i.e., context-dependent/partial/intent-implicit). Incomplete questions require context from previous questions and answers in the conversation to be answered correctly.

\myparagraph{Answer} An answer $a$ to $q$ is a crisp phrase (or a list) that satisfies the intent in $q$. In a heterogeneous scenario, the answer phrase $a$ can be an entity or literal (constant) coming out of the KB or a table or infobox, or any span of short text from the document corpus.

\myparagraph{Conversation} A conversation $C$ consists of a sequence of questions ($q^0, q^1, \ldots$) and corresponding answers ($a^0, a^1, \ldots$) (see Sec.~\ref{sec:intro} for an example). The first question $q^0$ in $C$ is complete, while follow-up questions are usually incomplete.

\myparagraph{Turn} A turn in $C$ consists of a specific $\langle q^i, a^i \rangle$ pair. For example, the second turn refers to $\langle q^1, a^1 \rangle$.

\myparagraph{Knowledge base} A knowledge base is a set of facts, where each fact is 
a \struct{<subject, predicate, object>} (SPO) triple, optionally augmented by \struct{<qualifier predicate, qualifier object>} pairs which specify additional information for the main triple (e.g. <\struct{Game of Thrones, cast member, Nikolaj Coster-Waldau; character role, Jaime Lannister}>). Subjects are entities (\struct{Game of Thrones)}, while objects can be entities, types (\struct{human}) or literals (constants such as numbers with or without units, dates like \struct{11 June 1969}, etc.). Predicates (\struct{cast member}) denote relationships.

\myparagraph{Text collection} A text collection is a set of documents, where each document consists of a sequence of sentences.

\myparagraph{Table} A table is a structured relational construct consisting of cells organized into rows and columns, with optional row and column headers. Cell values are typically entities or literals, while headers are often predicates.

\myparagraph{Infobox} An infobox is a list of salient attribute-value pairs about an entity. A Wikipedia infobox appears on the top right corner of the entity's Wikipedia page. Infobox entries resemble KB-facts, but they are not necessarily clean in terms of entity linkage (e.g., a birthplace could be given as a string with city, country or other regional variations).

\myparagraph{Evidence} An evidence 
is a unit of retrieval that can come from any of the heterogeneous sources above: a KB-fact, a text-sentence, a table-record (row), or an infobox-entry.
Evidences form the sources of answer candidates.

\myparagraph{Answering evidence} An answering evidence is an evidence which
contains
at least one correct answer $a$.
It can either mention the answer as a string,
or include an answer entity (in case of KB-facts).

%% file: sections/03-method.tex
\section{The \convinse Method}
\label{sec:method}

\begin{figure} [t]
     \includegraphics[width=\columnwidth]{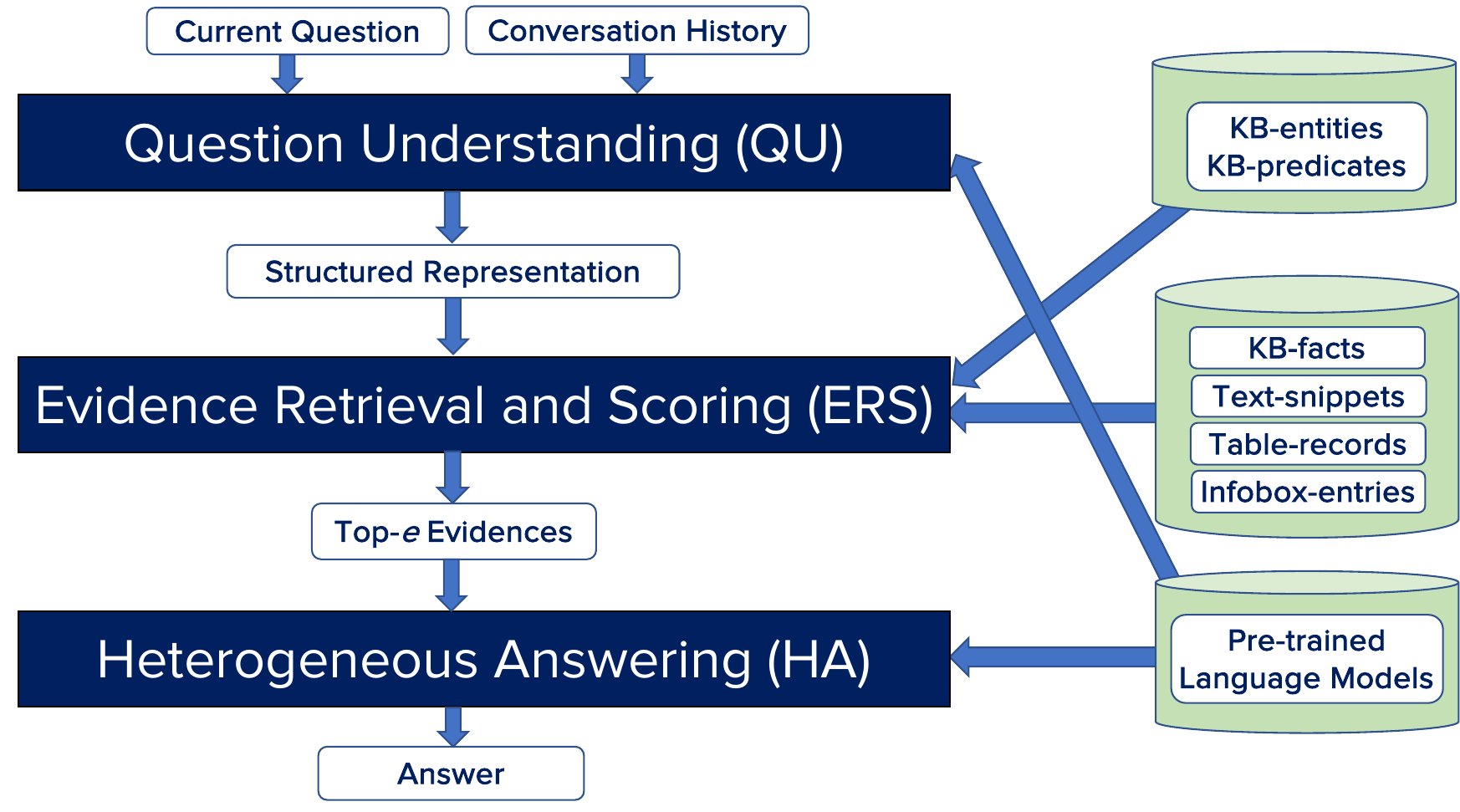}
     \caption{An overview of the \convinse system.}
     \label{fig:overview}    
     \vspace*{-0.2cm}
\end{figure}

Fig.~\ref{fig:overview} shows an overview of the \convinse architecture. The following subsections discuss the three steps: question understanding, evidence retrieval and scoring, and heterogeneous answering. 

\subsection{Question understanding (QU)}
\label{subsec:qu}

Follow-up questions in a conversation ($q^1, q^2, \ldots$) usually contain implicit intent. 
A key challenge in ConvQA, therefore, is to understand the flow of the conversation, 
towards deriving intent-explicit structured representations (SR) of the user's information need in $q^i$.
Instead of trying to generate a full-fledged question, we rather aim
to capture the semantic structure, using
a \textit{syntax-agnostic} approach.
This can be perceived as a ``logical form'' for heterogeneous QA,
where no explicit grounding or canonicalization is possible.
The representation is purely on the question-level,
and thus agnostic to the information sources that are used during the answering process.
However, it can readily be matched with different kinds of evidences, which often take the form of keyword phrases (e.g. text snippets or verbalized table records).
The SR is loosely inspired by work on quantity queries~\cite{ho2019qsearch}.

Specifically, an SR is a 4-tuple holding a slot for each of:
\squishlist
\item Context entities (depicted in {\color{gray} gray} in Table~1), 
\item Question entities (in {\color{red} red}), 
\item Question 
predicates (in {\color{blue} blue}), and,
\item Expected answer types (in {\color{cyan} cyan}).
\squishend

\myparagraph{Context and question entities}
As an example, consider the gold SR for $q^1$ of the running example:
$\langle$ \struct{\textcolor{gray}{GoT} | \textcolor{red}{the dwarf} | \textcolor{blue}{who played} | \textcolor{cyan}{human}} $\rangle$.
The context entity (\textcolor{gray}{GoT}  in this case) is an entity mention from the conversational context.
The question entity is the entity mention targeted in the current question (e.g. \struct{\textcolor{red}{the dwarf}}).
Here, the context entity makes the question entity explicit, indicating that the question is on the dwarf in Game of Thrones.
Inferring the question entity may need to take the history into account (e.g., for $q^3$ in Table~\ref{tab:qu}).
The context entity and question entity
can consist of multiple such mentions.
This is required for questions such as \phrase{Where did Dany and Jon first meet?},
with the gold SR being $\langle$ \struct{\textcolor{gray}{GoT} | \textcolor{red}{Dany and Jon Snow} | \textcolor{blue}{first meet} | \textcolor{cyan}{location}} $\rangle$.

\myparagraph{Question predicates}
The question predicate 
is the counterpart to the relation or attribute of interest in a logical form. 
However, it is merely a surface phrase, 
without any normalization or mapping to a KB.
This way, it is easy to match it against any kind of information source.
For example, the question predicates \struct{\textcolor{blue}{who played}} or \struct{\textcolor{blue}{first meet}} can be matched with
evidences from KB or text-snippets from documents or
table headers, alike.

\myparagraph{Answer types}
Expected answer types assist the answering model in detecting and eliminating spurious answer candidates~\cite{ziegler2017efficiency,saharoy2021question}.
In general, multiple types could be inferred here.
The question predicate \struct{\textcolor{blue}{first meet}} alone could imply the answer type to be either \phrase{date} or \phrase{location}.
Stopwords like \phrase{where} are often disregarded by downstream QA models~\cite{christmann2019look};
in contrast, 
the SR answer type retains this information and would infer only the correct \phrase{location}.
Further, this type can help in identifying the expected granularity of the answer.
For the question \phrase{When is his birthdate?}, one would expect a complete date with day, month and year as the answer,
but for
\phrase{When did they win their last world cup?} the corresponding year would be enough and actually desired.
\phrase{Date} and \phrase{year} would respectively populate the fourth slot in these cases.

Specific slots in the SR can  be left blank.
For $q_4$, the question entity \struct{\textcolor{gray}{GoT}} is already explicit, and thus no context entity is required: $\langle$ \struct{\textcolor{gray}{\_} | \textcolor{red}{GoT} | \textcolor{blue}{duration of an episode} | \textcolor{cyan}{number}} $\rangle$.

The SR generation is implemented by fine-tuning a pre-trained sequence generation model. 
We tried BART~\cite{lewis2020bart} and T5~\cite{raffel2020exploring} in preliminary experiments,
and found BART to perform
better.
BART is particularly effective when information is copied and manipulated from the input to generate the output autoregressively~\cite{lewis2020bart}, which is exactly the setting here.
The conversation history and the current question concatenated with a delimiter constitute the input,
and the SR is the output.
When encoding the history and the current question, the model 
considers cross-attention between turns, identifying
relevant parts from the conversation history.

\begin{figure} [t]
     \includegraphics[width=\columnwidth]{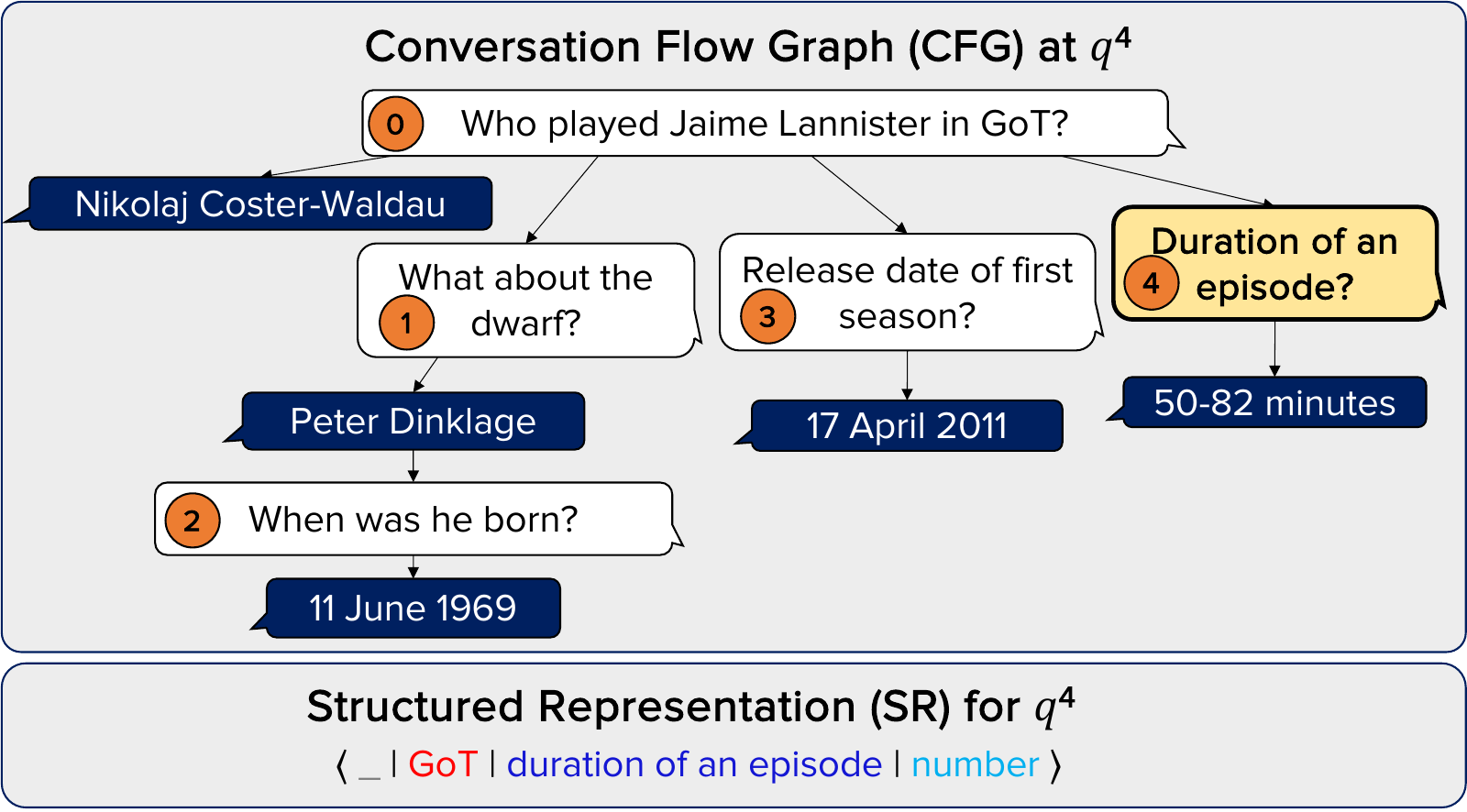}
     \caption{Conversation flow graph for our running example.}
     \label{fig:flow}
     \vspace*{-0.1cm}
\end{figure}

\myparagraph{SRs and explainability} One of our primary goals in \convinse was to produce intermediate representations for end users as we proceed through the QA pipeline.
Concretely, understanding the flow within the conversation is an essential problem in ConvQA~\cite{christmann2019look, huang2018flowqa}.
While the SR itself is human-readable, when presented only with the SR (or some rewritten/resolved question), certain decisions
of the ConvQA system might not be immediately obvious to a real user.
Here, we propose an intuitive
mechanism to infer and present the conversational flow to a user:
given the generated SR, we identify the source turn for each word, using exact match in the history, and consider such source turns as relevant for the current question. 
If there is no source turn, we consider the question at hand as self-sufficient.
A conversational flow graph (CFG) is established as follows:
questions and answers are nodes, and an edge connects a question to its relevant history. 
Due to the potential dependence of a turn on multiple preceding ones, the CFG for a conversation may not strictly be a tree, but rather a directed acyclic graph (DAG).
The CFG can be presented to the interested end user together with the SR, as depicted for $q^4$ in Fig.~\ref{fig:flow},
either for gaining confidence in final answers, or for scrutinizing error cases. 

\subsection{Evidence retrieval and scoring (ERS)}
\label{subsec:ers}

\begin{figure} [t]
     \includegraphics[width=\columnwidth]{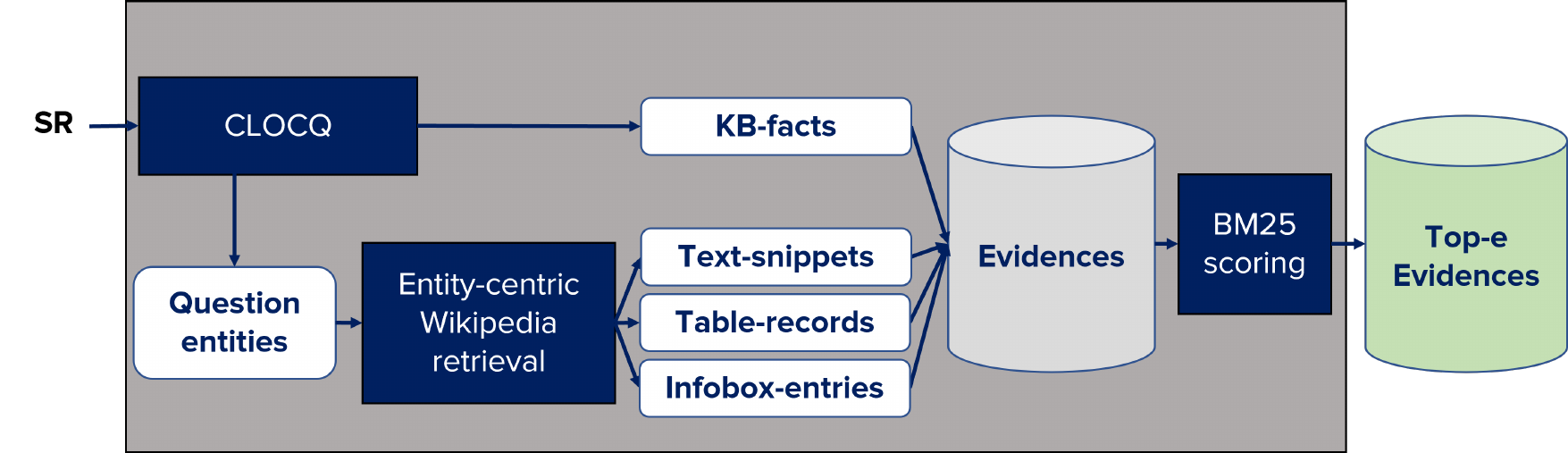}
     \caption{An overview of the ERS stage.}
     \label{fig:ers-overview}    
\end{figure}

\myparagraph{Evidence retrieval} At this stage, the goal is to retrieve relevant evidences given the generated SR.
We convert retrieved evidences on-the-fly
to verbalized NL forms~\cite{oguz2021unikqa}.
This is done for harnessing the state-of-the-art fusion-in-decoder (FiD) model downstream in our pipeline's final step (Sec.~\ref{subsec:fid}): FiD generates answers, given the question and a number of NL sentences as input.
Table~\ref{tab:evidences} shows example evidences from different sources for
the context entity \struct{Game of Thrones}.

For evidences from the \textbf{KB}, we make use of \clocq~\cite{christmann2022beyond} (available publicly at \url{https://clocq.mpi-inf.mpg.de/}), which is a recent method for retrieving relevant KB-facts and providing top-$k$ disambiguations (mappings to KB-items) for relevant cue words (entity, predicate, and type mentions), given an explicit user question.
Since questions are treated as keyword queries, the SR can directly be fed into \clocq (removing the separator `|' during retrieval).
KB-facts are verbalized~\cite{oguz2021unikqa,jia21complex,pramanik2021uniqorn}, separating the individual constituents of a fact by a comma.
These mappings
between KB-item mentions in verbalized facts
and KB-item IDs
are retained in-memory
to help us later during evaluation (Sec.~\ref{subsec:metric}).

For \textbf{text-corpus} evidences, we take an entity-centric view of Wikipedia:
for each of the disambiguated entities from \clocq, we retrieve the corresponding Wikipedia page.
For example, for the SR $\langle$ \struct{\textcolor{gray}{GoT} | \textcolor{red}{Jaime Lannister} | \textcolor{blue}{who played} | \textcolor{cyan}{human}} $\rangle$, we would consider the Wikipedia pages for \struct{Jaime Lannister}, \struct{Game of Thrones (TV Series)}, \struct{Game of Thrones (A Song of Ice and Fire)},
and so on (\clocq allows for multiple disambiguations per question token).
The text within the page is split into single sentences.

We extract
\textbf{tables and infoboxes}
from the retrieved Wikipedia pages.
Every table row is transformed to text individually,
concatenating
cell values with the respective column headers with an \phrase{is} in between, and separating such $\langle$cell, header$\rangle$ pairs with a comma~\cite{oguz2021unikqa}.
For each infobox attribute (similar to predicates in a KB), we concatenate the corresponding lines, having again, a comma as separator.
The page title is prepended to all evidences from Wikipedia as additional context.

In addition, we
exploit 
anchor
links in evidences to other Wikipedia pages 
to map the corresponding entity mentions
(anchor texts) to KB-items.
Wikipedia links are transformed to Wikidata IDs using a dictionary.
Similarly, dates and years in evidences are identified and normalized to the standard KB format using simple patterns.
We keep such $\langle$entity mention, KB-item$\rangle$ pairs for all evidences in memory, 
as 
this helps us later in grounding answers to the KB during evaluation (Sec.~\ref{subsec:metric}).
For example, for the text evidence in Table~\ref{tab:evidences},
\phrase{Tyrion} might link to the Wikipedia page of \struct{Tyrion Lannister}: so we add the pair
$\langle$\phrase{Tyrion}, \struct{Tyrion Lannister}$\rangle$
as metadata to the corresponding evidence.

\myparagraph{Evidence scoring}
The set of evidences compiled by the previous stage can be quite large (several thousands),
which can affect the efficiency and effectiveness of the answering phase.
Therefore, we first reduce this set, keeping only the most relevant information.
Since all evidences are verbalized at this stage, each individual evidence
can be treated as a document, with the SR as the query.
We then use the standard IR model BM25~\cite{robertson2009probabilistic}
for retrieving the top-$e$ relevant pieces of information.

\subsection{Heterogeneous answering (HA)}
\label{subsec:fid}
Given the top-$e$ relevant evidences per question, we make use of a state-of-the-art fusion-in-decoder~\cite{izacard2021leveraging} (FiD) model.
Different from the typical span prediction in the retrieve-and-read paradigm~\cite{chen2017reading},
FiD \textit{generates} the answer, following a sequence-to-sequence approach.
FiD first pairs every evidence with the question, which is the SR in our case,
and encodes these pairs, leveraging cross-attention to identify relevant information.
The concatenation of these encodings is then fed into
a decoder, which generates the answer autoregressively.

\subsection{Distantly supervised labeling}
\label{subsec:dist}

\myparagraph{Intuition} To train the QU phase of \convinse,
we need a collection of $\langle$history, question$\rangle$-pairs,
with the corresponding gold SRs.
Annotating conversations with such information
is tricky with crowdworkers, and expensive too
(c.f. Sec.~\ref{sec:data}).
Even the annotation of conversations with completed questions is
harder and much more expensive
than collecting plain sequences of question-answer pairs.
Therefore, we devise a mechanism to automatically generate the gold SRs from pure conversations.
Our technique is based 
on the following intuition:
if a piece of information (e.g. an entity or relation phrase) in previous turns is essential for 
the understanding of the current incomplete question
and
this information has been left implicit by the user,
then
it should be included in a completed version of the question.
Consider this example:
\begin{Snugshade}
    \utterance{$q^0$: Who played Jaime Lannister in GoT?}\\
    \indent $a^0$: \struct{Nikolaj Coster-Waldau}\\
    \indent \utterance{$q^1$: What about the dwarf?}
\end{Snugshade}
\noindent It is 
unlikely that proper evidences (on Game of Thrones, given $q^0$), can be found
for the incomplete question 
$q^1$.
However, once \phrase{GoT} is added to $q^1$ (e.g., \phrase{What about the dwarf in GoT?}),
\textit{answering evidences} (defined in Sec.~\ref{sec:concepts})
can be retrieved.
This suggests that
the phrase \phrase{GoT} should feature in the SR for $q^1$.

\myparagraph{Implementation}
Based on this idea, we create data for training the QU model as follows:
starting with the complete question $q^0$, we retrieve all evidences
(Sec.~\ref{subsec:ers}).
Since our retrieval is entity-centric, we can identify entity mentions which bring in \textit{answering evidences}:
for each evidence, the retriever returns the text span from the input question
that
the evidence
was retrieved for.
Such entity mentions are considered relevant for the respective conversation turn.
For the incomplete follow-up questions, we iteratively add such relevant entity mentions from previous turns, and then
retrieve evidences for the current question at hand.
When adding the entity mention results in answering evidences being retrieved,
we consider the entity as relevant for the current turn.
Similarly, entity mentions in the current turn are identified as relevant, if answering evidences are retrieved for them.

The gold SR is then constructed heuristically:
i) if there are relevant entities from the current turn, then these feature in the question entity slot,
and relevant entities (if any) from previous turns become context entities in the SR, and,
ii) if there are only relevant entities from previous turns, then these become question entities.
The remaining words in the current question (except for stopwords) fill up the question predicate slot.
The expected answer type is directly looked up from the KB, using the gold answer.
Since the KB may have several types for the same entity, we take the most frequent one
to avoid very specific types. For example, \struct{Tyrion Lannister} has types \struct{fictional human} (more frequent) and \struct{GoT character} (less frequent) in Wikidata: so we only consider \struct{fictional human} in our SR.

%

\subsection{Training the \convinse framework}
\label{subsec:training}

We train the QU model first, using data generated in the manner discussed in Sec.~\ref{subsec:dist}.
After training, we directly apply the QU model on the whole data (including training data),
to generate SRs.
In the remainder of the approach, we utilize only the SRs generated by our model.
When training the FiD model, we skip instances for which the top-$e$ evidences
do not have the answer, since the model would not be able to find the answer
within the evidences, and could
hallucinate at inference time~\cite{roller2021recipes}.
This way, the model is taught to predict the answer from the input.
Since we would like to treat evidences from different sources in a unified way,
only one model is trained on the top-$e$ evidences after retrieval from all sources.
To demonstrate the robustness of \convinse, this same model is subsequently used for all combinations of input sources, including
inputs from single sources.

\begin{table} [t] 
    \caption{Verbalized evidences from different input sources.}
    \vspace*{-0.3cm}
    \resizebox*{\columnwidth}{!}{
	\begin{tabular}{p{1.0cm} p{6.5cm}}
	    \toprule
	    \textbf{KB}             &  \utterance{Game of Thrones, cast member, Nikolaj Coster-Waldau, character role, Jaime Lannister}   \\
	    \textbf{Text}           &  \utterance{Game of Thrones, The third and youngest Lannister sibling is the dwarf Tyrion (Peter Dinklage) (...).} \\
	    \textbf{Table}          &  \utterance{Game of Thrones, Season is Season 1, (...), First aired is April 17, 2011 (...).}\\
	    \textbf{Infobox}        &  \utterance{Game of Thrones, Running time, 50–82 minutes}\\
	    \bottomrule
	\end{tabular} }
	\label{tab:evidences}
\end{table}

%% file: sections/04-benchmark.tex

\section{The \convmix benchmark}
\label{sec:data}

\begin{table} [t] \small
	\centering
	\caption{Basic statistics for the \convmix benchmark.}
	 \vspace*{-0.3cm}
	\resizebox*{\columnwidth}{!}{
	\begin{tabular}{p{2.2cm} p{5.5cm}}
		\toprule
		Title			        & Generate 5 conversations for question answering	\\
		Description		        & Choose entities of your choice from five domains, generate questions about them, and find answers from Wikidata and Wikipedia\\ \midrule
		Participants			& $32$ unique Master Turkers \\
		Time allotted (HIT)	    & $4$ hours maximum \\ 
		Time taken (HIT) 	    & $1.5$ hours on average \\ 
		Payment per HIT			& $15$ USD \\ 
		Domains                 & Books, Movies, Music, TV series, Soccer           \\
		Conversations           & $3000$ \\
	    Questions               & $16000$  \\
	    Question length         & $8.78$ words (initial), $5.19$ (follow-ups), $5.87$ (all) \\ 	    
	    Answer size             & $1.02$ entities/strings on average \\ 
	    Entities covered        & $5418$ (long-tail: $2511$, with <$50$ KB-facts) \\ 
	    Heterogeneity           & $2626$ conversations (>$1$ source used by Turker) \\ \bottomrule
	\end{tabular}}	
	\label{tab:amt}
\end{table}

\myparagraph{Limitations of existing ConvQA benchmarks} Notable efforts at ConvQA benchmarking like \textsc{QuAC}~\cite{choi2018quac} (text), \textsc{CoQA}~\cite{reddy2019coqa} (text), \textsc{SQA}~\cite{iyyer2017search} (tables), \textsc{ConvQuestions}~\cite{christmann2019look} (KB), and \textsc{CSQA}~\cite{saha2018complex} (KB) assume a single answering source. Rather than the easier option of artificially augmenting any of these with heterogeneous inputs,
we believe that it is much more natural and worthwhile to build a new resource from scratch by users browsing through a mixture of sources, as they would do in a typical information seeking session on the Web.

\myparagraph{Initiating conversations} How a user initiates a conversation is a key conceptual challenge that needs to be overcome in creating good ConvQA benchmarks. One could, for example, provide users with passages or documents, and ask them to create a sequence of questions from there~\cite{choi2018quac,reddy2019coqa}. Alternatively, one could also provide annotators with some conversation from a benchmark so far, and request their continuation in some fashion~\cite{kaiser2021reinforcement}. Large-scale synthetic benchmarks would try to automate this as far as possible using rules and templates~\cite{saha2018complex}. In keeping with our philosophy of natural conversations, we asked users to start with an entity of their choice (instead of spoonfeeding them with one, which could be counterproductive if the user has no interest or knowledge about the provided entity). Real conversations between humans, or several search sessions, often start when users have queries about such seed or topical entities.
With the first question initiating the conversation, we collected four follow-up questions (total of five turns) that build upon the ongoing conversational context.

\myparagraph{Quality control} The study was conducted on the popular crowdsourcing platform Amazon Mechanical Turk (AMT), where we allowed only Master Workers
to participate, for quality assurance. We also blocked single and sets of annotators who demonstrated evidence of excessive repetition or collusion in their annotations. Since the task is non-trivial for an average Turker (requires understanding of factoid questions, and familiarity of knowledge sources like Wikidata and Wikipedia, along with entities and literal answers), we also included quite a few internal checks and warnings that could prompt users for unintentional mistakes before task submission.
Guidelines are given as text, but we also provided a video illustrating the conversation generation process by some examples.
Workers with notably diverse and interesting conversations were awarded with a 5 USD bonus.
Interestingly, several Turkers providing high-quality conversations seemingly found the task engaging (we provided a free-text feedback box), and submitted more than $20$ HITs.
The authors conducted semi-automatic post-processing, validation and cleaning of the benchmark. Several issues were also resolved by meticulous manual inspection.
For example, we ran \clocq~\cite{christmann2022beyond} on the initial questions, and manually inspected cases for which no answering evidences were found, to identify and rectify cases in which the initial questions themselves were unanswerable.
Such cases are specifically problematic, because the whole conversation might become unanswerable.

\myparagraph{Ensuring heterogeneity} Last but not the least, ensuring answer coverage over heterogeneous sources was a key concern. Here, we again kept it natural, and encouraged users not to forcibly stick to any particular source during their conversation. Interestingly, out of $3000$ conversations, only $374$ used exactly one source. A majority ($1280$) touched three sources, $572$ touched four, while $774$ used two inputs. Finally, note that this is only the source that the annotator used during her search process: it is quite possible that the answer can be located in other
information sources
(see the field $[\cdot]$ below answers in Table~\ref{tab:convmix}), thereby enabling future benchmark users to exploit answer redundancy.

\myparagraph{Collecting longer conversations}
We initially collected
$2800$ conversations
with 
five turns (referred to as \convmix-5T).
However, there can also be cases in which users wish to dive deep into a specific topic,
or other curiosities
arise
as the conversation continues.
In such situations, conversations can easily go beyond five turns, making the understanding of
the conversational flow even more challenging for the ConvQA system.
Therefore, we collected $200$ additional conversations with ten turns (denoted \convmix-10T), to test the generalizability
of ConvQA systems over longer conversations.
On manual investigation, we found that there are naturally more topic drifts within these conversations.
These $2$k ($200 \times 10$) questions are only used
as an additional test set
to serve as a robustness check for pre-trained models.
Thus, our complete benchmark \convmix (3000 conversations in total) is made up of subsets \convmix-5T (2800 conversations, 5 turns each)
and
\convmix-10T (200 conversations, 10 turns each).

\myparagraph{Collected fields} We collected the following annotations
from crowdworkers:
i) conversational questions, ii) intent-explicit versions of follow-up questions,
iii) gold answers
as plain texts and Wikidata URLs,
iv) question entities,
v) question paraphrases, and
vi) sources used for answer retrieval. We believe that this additional metadata
will make our resource useful beyond QA (in question rewriting and paraphrasing, for example).
Most questions had exactly one correct answer, with the maximum being six. Table~\ref{tab:amt} summarizes notable properties of our study and benchmark, while Table~\ref{tab:convmix} reports interesting representative examples.
Note that HIT specific entries, like the payment per HIT, are given for the collection of conversations with five turns.
The respective numbers were doubled for the collection of conversations with ten turns.

\begin{table*} [t] \small
	\centering
	\caption{Representative conversations in \convmix. The types of sources which can be used for answering are given in brackets.}
    \vspace*{-0.3cm}
	\resizebox*{\textwidth}{!}{
		\begin{tabular}{p{0.5cm} | p{3cm} | p{3cm} | p{3cm} | p{3cm} | p{3cm}}
			\toprule
			\textbf{Turn}	& \textbf{Books}	&	\textbf{Movies}		&	\textbf{Music}	&	\textbf{TV series} &	\textbf{Soccer} \\ \toprule
			$q^0$			& \utterance{Who wrote Slaughterhouse-Five?}	&	\utterance{Who played Ron in the Harry Potter movies?}	&	\utterance{What was the last album recorded by the Beatles?}	&	\utterance{Who is the actor of Rick Grimes in The Walking Dead?}	&	\utterance{Which national team does Kylian Mbappé play soccer for?} \\
			$a^0$			& \struct{Kurt Vonnegut}	&	\struct{Rupert Grint}	&	\struct{Let It Be}	&	\struct{Andrew Lincoln}	&	\struct{France football team} \\
			    	        & [KB, Text, Info]	&	[KB, Text]	&		[KB, Text, Table]	&		[KB, Text, Table]	&		[KB, Text, Info, Table] \\\midrule
			$q^1$			& \utterance{Which war is discussed in the book?}	&	\utterance{Who played Dumbledore?}	&	\utterance{Where was their last paying concert held?}	&	\utterance{What about Daryl Dixon?} &	\utterance{How many goals did he score for his home country in 2018?} \\
			$a^1$			& \struct{World War II}	&	\struct{R. Harris, M. Gambon}	&	\struct{Candlestick Park}	&	\struct{Norman Reedus}	&	\struct{9} \\ 
			                 & [KB, Text]	&	[Text, Table]	&		[Text]	&		[KB, Text, Table]	&		[Table] \\\midrule
			$q^2$			& \utterance{What year was it's first film adaptation released?}	&	\utterance{What's the run time for all the movies combined?}	&	\utterance{What year did they break up?}	&	\utterance{did he also play in Saturday night live?} &	\utterance{place of his birth?}  \\
			$a^2$			& \struct{1972}	&	\struct{1179 minutes}	&	\struct{1970}	&	\struct{Yes}	&	\struct{Paris} \\ 
			                & [KB, Text, Table, Info]	&	[KB, Info]	&		[KB, Text, Info]	&		[Text]	&		[KB, Text, Info] \\\midrule
			$q^3$			& \utterance{Who directed it?}	&	\utterance{Who was the production designer for the films?}	&	\utterance{Who was their manager?}	&	\utterance{whom did he play?} &	\utterance{award he got in 2017?"} \\
			$a^3$			& \struct{George Roy Hill}	&	\struct{Stuart Craig}	&	\struct{Brian Epstein}	&	\struct{Daryl Dixon}	&	\struct{Golden Boy} \\ 
			                & [KB, Text, Table, Info]	&	[KB, Text, Table]	&		[KB, Text]	&		[Text]	&		[KB, Table] \\\midrule
			$q^4$			& \utterance{What was the final film that he made?}	&	\utterance{Which movie did he win an award for working on in 1980?}	&	\utterance{What was their nickname?}	&	\utterance{production company of the series?} &	\utterance{Who is the award conferred by?} \\ 
			$a^4$			& \struct{Funny Farm}	&	\struct{The Elephant Man}	&	\struct{Fab Four}	&	\struct{NBC Studios}	&	\struct{Tuttosport} \\ 
			                & [KB, Text, Table]	&	[Text]	&		[KB, Text]	&		[KB, Text, Info]	&		[KB, Text, Info] \\   \bottomrule
	\end{tabular}}
	\label{tab:convmix}
	\vspace*{-0.1cm}
\end{table*}

%% file: sections/05-experiments.tex
\section{Experimental setup}
\label{sec:exp-setup}

We conduct all experiments on the \convmix benchmark.
We split the part of \convmix with five turns, \convmix-5T, into train, development and test sets with the ratio 60:20:20.
\convmix-10T is 
used only as a separate test set.

\subsection{Heterogeneous sources}
\label{subsec:sources}

\convinse and all baselines run on the same data collections. As our knowledge base, we take the 31 January 2022 complete NTriples dump\footnote{\url{https://dumps.wikimedia.org/wikidatawiki/entities/}} of Wikidata, one of the largest and best curated KBs today. It consists of about $17$B triples, consuming about $2$ TB disk space. We access the KB via the recently proposed \clocq~\cite{christmann2022beyond} interface, that reduces the memory overhead and efficiently returns KB-facts for queried entities.
The text collection is 
chosen to be the
English Wikipedia (April 2022).
The benchmark-relevant subset of Wikipedia is comprised of the pages of entities detected via \clocq. 
Documents are split into sentences using spaCy\footnote{https://spacy.io/api/sentencizer}.
All tables and infoboxes originating from the retrieved Wikipedia pages together constitute the respective answering sources.
We parse Wikipedia tables using WikiTables\footnote{https://github.com/bcicen/wikitables}, and concatenate entries in the obtained JSON-dictionary for verbalization.
This procedure also includes conversions for tables with nested structure.
Infoboxes are detected using Beautiful Soup\footnote{https://beautiful-soup-4.readthedocs.io/en/latest/}.

\subsection{Baselines}
\label{subsec:baselines}

There are no prior works for ConvQA over heterogeneous sources. Thus, to compare the proposed \convinse pipeline with alternative choices, we adapt state-of-the-art question understanding (in this case, rewriting and resolution) methods from the IR and NLP literature.
These serve as competitors for our SR generation phase. We then provide these baselines with exactly the same ERS and HA phases that \convinse has, to complete end-to-end QA pipelines.

\myparagraph{Prepending history turns} Adding  turns from the history to the beginning of the current question is still considered
a simple yet tough-to-beat baseline in almost all ConvQA tasks~\cite{qu2019bert,christmann2019look,vakulenko2021question,kaiser2021reinforcement},
and so we investigate the same here as well. Specifically, we consider four variants: i) add only the initial turn $\langle q^0, a^0\rangle$, as it often establishes the topic of the conversation (\textbf{Prepend init}); ii) add only the previous turn $\langle q^{i-1}, a^{i-1}\rangle$, as it sets immediate context for the current information need (\textbf{Prepend prev}); iii) add both initial and previous turns (\textbf{Prepend init+prev}); and iv) add all turns $\{\langle q^t, a^t\rangle\}_{t = 0}^{i-1}$ (\textbf{Prepend all}).

\myparagraph{Question rewriting} We choose a very recent T5-based rewriting model~\cite{raposo2022question}. The method
is trained on the \textsc{Canard} question rewriting benchmark~\cite{elgohary2019can}.
The model is fine-tuned on \convmix, using the $\langle$full history, current question$\rangle$-pairs as input, and the respective completed questions (available in the benchmark) as the gold label.

\myparagraph{Question resolution} We use \quretec~\cite{voskarides2020query} as a question resolution baseline, treating context disambiguation as a term classification problem.
A BERT-encoder is augmented with a term classification head,
and predicts for each history term whether the word should
be added to the current question.
The same distant supervision strategy (Sec.~\ref{subsec:dist}) as used by \convinse is employed for generating annotated data for \textsc{QuReTeC} (trained on \convmix).

\subsection{Metrics}
\label{subsec:metric}

\myparagraph{Measuring retrieval effectiveness} To evaluate retrieval quality, we use \textbf{answer presence} as our metric. It is a binary measure of whether one of the gold answers is present in the top-$e$ evidences ($e = 100$ in all experiments).

\myparagraph{Measuring answering effectiveness} We use one of the standard metrics for factoid QA~\cite{saharoy2021question}, \textbf{precision@1 (P@1)}, since FiD generates a unique answer.
FiD generates plain strings as answers:
evaluation for such strings with exact match for computing P@1 can often be problematic~\cite{si2021whats},
since the correct answer could be expressed in different ways (e.g. 
\{\phrase{Eddard Stark}, \phrase{Ned Stark}\}, or
\{\phrase{11 June 1969}, \phrase{11-06-1969}, \phrase{June 11, 1969}\}).
Therefore, we try to normalize the answer to the KB, whenever possible, to allow for a fair comparison across systems.
We search through $\langle$entity mention, KB-item$\rangle$ pairs coming from the evidence retrieval phase (Sec.~\ref{subsec:ers}).
If there is a perfect match between the entity mention and the predicted answer string, we return the corresponding
KB-item as the answer.
If there is no such perfect match, we compute the Levenshtein distance~\cite{levenshtein1966binary} between
the predicted answer and entity mentions from $\langle$entity mention, KB-item$\rangle$ pairs.
The KB-item for the entity mention with the smallest edit distance is
used
as the answer in such cases. 
Note that such KB-items may also be normalized strings, dates, years, or numbers.
These normalized KB-items are compared to the gold answers in the benchmark for the computation of P@1.

\subsection{Configurations}
\label{subsec:init}

\convinse uses a fine-tuned BART-base model for generating structured representations. The default hyperparameters from the Hugging Face library were used\footnote{\url{https://huggingface.co/facebook/bart-base}}.
The maximum sequence length was set to $20$, and early stopping was enabled.
Three epochs were used during training, with a batch size of ten. $500$ warmup steps with a weight decay of $0.01$ turned out to be the most effective.
\clocq, that was used for evidence retrieval inside \convinse and all baselines, has two parameters: $k$ (number of disambiguations to consider for each question word), and $p$ (a pruning threshold). In this paper, we set $k$ = \struct{Auto} (\clocq dynamically sets the number of disambiguations), and $p = 1000$, as these performed the best on our dev set.
We used a Python implementation of BM25, with default parameters\footnote{\url{https://pypi.org/project/rank-bm25/}}.
Code for FiD is publicly available\footnote{\url{https://github.com/facebookresearch/FiD}}. 
FiD was trained on \convmix for its use in this work.
The number of input passages ($e$ in this paper) was retained at $100$ as in the original work~\cite{izacard2021leveraging}.
The maximum length of an answer was set to $10$ words. 
A learning rate of $5 \times 10^{-5}$ really proved effective, with a weight decay of $0.01$ and
AdamW as the optimizer.
All systems were trained on the \convmix train set, and all hyperparameters were tuned on the dev set.
All code was
scripted
using \texttt{Python}, making use of the popular PyTorch library\footnote{\url{https://pytorch.org}}.
Whenever a neural model was used, code was run on a GPU (single GPU, NVIDIA Quadro RTX 8000, 48 GB GDDR6).

%% file: sections/06-results.tex

\section{Results and insights}
\label{sec:res}

\begin{table*}
    \caption{Comparison of answer presence within top-100 retrieved evidences after QU + ER on the \convmix test set.} 
        \vspace*{-0.3cm}
    \newcolumntype{G}{>{\columncolor [gray] {0.90}}c}
    \newcolumntype{L}{>{}c}
      \resizebox*{\textwidth}{!}{
    	\begin{tabular}{l G G G G c c c c c c G}
    		\toprule
    		    \textbf{QU + ERS Method}	   & \textbf{KB} & \textbf{Text} & \textbf{Table} & \textbf{Info} & \textbf{KB+Text}  & \textbf{KB+Table} & \textbf{KB+Info} & \textbf{Text+Table}  & \textbf{Text+Info} & \textbf{Table+Info} & \textbf{All} \\
    		    \midrule
                \textbf{Prepend init + BM25} 
        		&	$0.380$	 &	$0.298$	 &	$0.120$ 	&	$0.331$ 	&	$0.415$  &	$0.386$ &	$0.406$  &	$0.297$	&	$0.329$  &	$0.331$	&	$0.419$  \\
        	   
        	    \textbf{Prepend prev + BM25}
        	    &	$0.342$	 &	$0.284$	 &	$0.095$ 	&	$0.295$ 	&	$0.382$ &	$0.347$ &	$0.372$ 	&	$0.284$ &	$0.317$   &	$0.306$	&	$0.392$  \\
        	   
        	    \textbf{Prepend init+prev + BM25}
        	    &	$0.440$  &	$0.366$   &	$0.137$ & 	$0.420$ & $0.486$	&	$0.443$ &	$0.479$ &	$0.359$  &	$0.407$  &	$0.409$ &	$0.495$    \\
        	    
        	    \textbf{Prepend all + BM25}
        	    &	$0.431$   &	$\textbf{0.367}$    &	$\textbf{0.148}$  &	$\textbf{0.430}$  &	$0.476$	 &	$0.437$  &	$0.468$  &	$\textbf{0.361}$ & 	$\textbf{0.411}$   &	$\textbf{0.419}$  &	$0.482$    \\ \midrule
        	    
        	    \textbf{Q. Resolution~\cite{voskarides2020query} + BM25 + FiD}
        		&	$0.414$  &	$0.311$   &	$0.115$ & 	$0.329$ & $0.445$	&	$0.419$ &	$0.437$ &	$0.312$  &	$0.356$  &	$0.341$ &	$0.453$    \\
        		
        		\textbf{Q. Rewriting~\cite{raposo2022question} + BM25 + FiD}
        		&	$0.434$  &	$0.315$   &	$0.114$ & 	$0.347$ & $0.460$	&	$0.435$ &	$0.461$ &	$0.319$  &	$0.362$  &	$0.336$ &	$0.465$    \\
        	    
        	   \midrule
        	    \textbf{\convinse (Proposed)} 
        	    &	$\textbf{0.475}$*  &	$0.352$   &	$0.117$ & 	$0.369$ & $\textbf{0.528}$*	&	$\textbf{0.486}$* &	$\textbf{0.507}$* &	$0.353$  &	$0.408$  &	$0.381$ &	$\textbf{0.542}$*    \\
        	   
    		\bottomrule
    	\end{tabular} 
    }
    \label{tab:ans-pres}
    \vspace{-0.3cm}
\end{table*}

\begin{table*} 
    \caption{Comparison of end-to-end (QU + ERS + HA) answering performance (P@1) on the \convmix test set.}
    \vspace*{-0.3cm}
    \newcolumntype{G}{>{\columncolor [gray] {0.90}}c}
    \resizebox*{\textwidth}{!}{
    	\begin{tabular}{l G G G G c c c c c c G}
    		\toprule
        		\textbf{QU + ERS + HA Method}	 & \textbf{KB}  & \textbf{Text}  & \textbf{Table}  & \textbf{Info} & \textbf{KB+Text} & \textbf{KB+Table} & \textbf{KB+Info} & \textbf{Text+Table} & \textbf{Text+Info}  & \textbf{Table+Info}  & \textbf{All}  \\ 
        		\midrule

                \textbf{Prepend init + BM25 + FiD} 
        		&	$0.211$  &	$0.174$  &	$0.065$  &	$0.200$   &	$0.246$ & 	$0.211$ &	$0.240$ &	$0.174$ &	$0.203$    &	$0.195$ &	$0.254$    \\
        		
        		\textbf{Prepend prev + BM25 + FiD} 
        		&	$0.179$  &	$0.190$  &	$0.052$  &	$0.212$   &	$0.238$ & 	$0.184$ &	$0.233$ &	$0.185$ &	$0.224$    &	$0.211$ &	$0.257$    \\
        		
        		\textbf{Prepend init+prev + BM25 + FiD} 
        		&	$0.234$  &	$0.233$  &	$\textbf{0.074}$  &	$0.276$   &	$0.290$  &	$0.238$ &	$0.292$  &	$\textbf{0.229}$ &	$0.274$  &	$\textbf{0.272}$ &	$0.312$    \\
        		
        		\textbf{Prepend all + BM25  + FiD}
        	    &	$0.230$  &	$\textbf{0.234}$   &	$\textbf{0.074}$ & 	$\textbf{0.282}$ & $0.290$	&	$0.238$ &	$0.282$ &	$0.224$  &	$0.267$  &	$0.265$ &	$0.300$    \\ \midrule
        		
        		\textbf{Q. Resolution~\cite{voskarides2020query} + BM25 + FiD}
        		&	$0.222$  &	$0.190$   &	$0.063$ & 	$0.219$ & $0.261$	&	$0.227$ &	$0.257$ &	$0.185$  &	$0.241$  &	$0.221$ &	$0.282$    \\
        		
        		\textbf{Q. Rewriting~\cite{raposo2022question} + BM25 + FiD}
        		&	$0.216$  &	$0.183$   &	$0.062$ & 	$0.219$ & $0.252$	&	$0.221$ &	$0.261$ &	$0.187$  &	$0.227$  &	$0.223$ &	$0.271$    \\
        		\midrule
        		\textbf{\convinse (Proposed)} &	$\textbf{0.251}$*	 &	$0.220$	 &	$0.062$ &	$0.258$	 &	$\textbf{0.317}$*	 &	$\textbf{0.257}$* &	$\textbf{0.310}$*	 &	$0.220$	 &	$\textbf{0.276}$ &	$0.253$	 &	$\textbf{0.342}$* \\
    		\bottomrule
    	\end{tabular}}
    \label{tab:qa-res}
    \vspace*{-0.4cm}
\end{table*}

We run \convinse and the baselines on the test set of \convmix (combined test sets of -5T and -10T), and report results in Tables~\ref{tab:ans-pres} and~\ref{tab:qa-res}. 
All metrics are micro-averaged over each conversational question that the systems handle, i.e. we measure the performance at a question-level.
Throughout this section, best performing variants in columns are marked in \textbf{bold}. An asterisk (*) denotes statistical significance of
\convinse
over the nearest baseline. The
McNemar's test was performed for binary variables like P@1, and the paired $t$-test otherwise, with $p < 0.05$.
All results are reported on the test set, except the ablation study, that was naturally conducted on the dev set.
If not stated otherwise, we make use of \textit{gold} answers for the previous turns in the conversation.
For example, for answering $q^4$ we assume gold answers $a^0$-$a^3$ to be known.

\subsection{Key findings}
\label{subsec:main-res}

\myparagraph{\convinse is viable for heterogeneous QA} The first and foremost takeaway 
is that our
proposed pipeline
is a viable approach for handling 
incomplete questions in conversations,
given heterogeneous
input
sources.
\convinse and most of the baselines
consistently reach $40$-$50\%$ on answer presence
in their evidence pools after QU and ERS,
with \convinse leading with $54\%$
(see Table \ref{tab:ans-pres}). 
These numbers set the upper bounds for
end-to-end QA after the answering phase,
which are in the ballpark of $25$-$34\%$
(see Table \ref{tab:qa-res}).

It is noteworthy that even the basic prepending baselines, despite generating fairly verbose question formulations ($15-26$ words, Table~\ref{tab:ques-len}) have very good answer presence in the top-100 evidences. This is largely due to the \clocq entity-based retrieval module, which
turned to be
quite
robust
even for
long queries.
Subsequently, BM25 scoring serves
as a necessary filter, to prune the evidence sets. Sizes of these evidence sets varied from $2.3$k evidences (\convinse) to $7.5$k (Prepend all), which
would have posed efficiency challenges to the final answering stage had the BM25 filter not been applied.

\myparagraph{\convinse outperforms baselines} We observe that the SRs in \convinse are significantly more effective than question rewriting/resolution and the prepend baselines. SRs provide the right balance between conciseness and coverage. Conciseness (SRs are just 6-7 words on average, Table~\ref{tab:ques-len}) helps effective retrieval with
the
IR model at the ERS stage, while expressive representation of the conversational context is crucial, too.
The expressivity in the SRs helps achieve the highest answer presence
after QU and ERS ($0.542$ in Table~\ref{tab:ans-pres}),
outperforming all baselines by substantial margins.
This advantage is carried forward to the answering stage
and results in the best end-to-end QA performance
($0.342$ in Table~\ref{tab:qa-res}).

The simple prepend baselines sometimes come surprisingly close, though, and perform better than the sophisticated rewriting or resolution methods.
Nevertheless, it is noteworthy that SRs and question rewriting/resolution have clear advantages in terms of interpretability. When a user wonders what went wrong upon receiving an incorrect (if not outright embarrassing) answer,
the system could show the CFG, SRs or the inferred complete questions as informative explanations -- helping the user to better understand and cope with the system's limitations and strengths.

Altogether, the absolute P@1 numbers still leave huge room for improvement. This shows that \convmix is indeed a challenging benchmark, with difficult user inputs where inferring the intent is hard, for example: \utterance{Which war is discussed in the book?} or \utterance{What was the final film he made?} (see Table~\ref{tab:convmix} for more examples).

\myparagraph{Combining heterogeneous sources helps} Another across-the-board observation is that combining knowledge sources is a strong asset for ConvQA.
Consider the values in the \phrase{All} columns
of Tables \ref{tab:ans-pres} and \ref{tab:qa-res}.
These numbers
are systematically and substantially higher than those in the columns for individual source types and even for pair-wise combinations.

To inspect whether these gains are not only from enhanced coverage,
but also leverage \textit{redundancy of information} across source types,
we measured the average P@1 for all
cases
where the questions had
1, 2, 3, or 4
evidences (among the top-$e$) containing the gold answer in the output of the ERS stage.
The P@1 improves steadily as the answer can be found several times, i.e. as information becomes redundant, being $0.428$ for one, 
$0.658$ for two, $0.713$ for three, and $0.763$ across instances with four answering evidences.
%

\myparagraph{\convinse excels in realistic setup with \emph{predicted answers}}
The experiments so far are conducted assuming the gold answers for the previous turns in the conversation to be given.
However, in a realistic setup, the ConvQA system would not know these answers.
Therefore, we conducted an additional experiment, in which we used the \textit{predicted answers} by the system for the previous turns, when generating the outputs of the QU phase.
The results of this experiment are shown in Table~\ref{tab:pred-answer}.
\convinse outperforms all methods significantly on conversations of length both five and ten, even though it has never seen such 10-length conversations during training.
Another observation is that the performance for the \phrase{Prepend all} baseline drops for \convmix-10T,
which might be due to the much longer QU outputs generated as the conversation continues longer.
In general, the performance of all methods drops a bit (about $0.057$ P@1 on average, c.f. Tables~\ref{tab:qa-res} and~\ref{tab:pred-answer})
when changing from gold answers to predicted answers.

\subsection{In-depth analysis}
\label{sec:analysis}


\begin{table} 
    \caption{P@1 of ConvQA systems when using the \emph{predicted answers} for the previous turns in the ongoing conversation.}
    \vspace*{-0.3cm}
    \newcolumntype{G}{>{\columncolor [gray] {0.90}}c}
    \resizebox*{\columnwidth}{!}{
    	\begin{tabular}{l c c c}
    		\toprule
    		    \textbf{Method}	            & \textbf{\convmix-5T}  & \textbf{\convmix-10T} & \textbf{\convmix} \\
    		     \textbf{(+ BM25 + FiD)}    & (2800 questions)  &   (2000 questions) & (4800 questions) \\
    		    \midrule
        	    \textbf{Prepend init}       &	$0.276$ &	$0.178$ &	$0.235$ \\
        	    \textbf{Prepend prev}       &	$0.190$ &	$0.123$ &	$0.162$ \\
        	    \textbf{Prepend init+prev}  &	$0.277$ &	$0.195$ &	$0.243$ \\
        	    \textbf{Prepend all}        &	$0.284$ &	$0.168$ &	$0.236$ \\ \midrule
        	    \textbf{Q. Resolution~\cite{voskarides2020query}}   &	$0.283$ &	$0.188$ &	$0.243$ \\
        	    \textbf{Q. Rewriting~\cite{raposo2022question}}    &	$0.258$ &	$0.168$ &	$0.221$ \\ \midrule 
        	    \textbf{\convinse}          &	$\textbf{0.321}$* &	$\textbf{0.217}$* &	$\textbf{0.278}$* \\
    		\bottomrule
    	\end{tabular} }
    \label{tab:pred-answer}
    \vspace*{-0.3cm}
\end{table}

\begin{table} 
    \caption{P@1 of ConvQA systems over turns.}
    \vspace*{-0.3cm}
    \newcolumntype{G}{>{\columncolor [gray] {0.90}}c}
    \resizebox*{\columnwidth}{!}{
    	\begin{tabular}{l G G c c c c}
    		\toprule
    		                                   & \multicolumn{2}{G}{\textbf{\convmix-5T}} 	 & \multicolumn{4}{c}{\textbf{\convmix-10T}} \\
    		    \textbf{Method (+ BM25 + FiD)} & \textbf{1}  & \textbf{2--5}  & \textbf{1}  & \textbf{2--4} & \textbf{5--7}  & \textbf{8--10} \\
    		    \midrule
        	    \textbf{Prepend init}                                   &	$0.388$ &	$0.271$ &	$0.320$  &	$0.235$ &	$0.185$   &	$0.128$  \\
        	    \textbf{Prepend prev}                                   &	$0.371$ &	$0.243$ &	$0.325$  &	$0.243$ &	$0.232$   &	$0.218$  \\
        	    \textbf{Prepend init+prev}                              &	$0.370$ &	$0.315$ &	$\textbf{0.335}$  &	$\textbf{0.310}$ &	$0.308$   &	$0.245$ \\
        	    \textbf{Prepend all}                                    &	$0.375$ &	$0.323$ &	$0.305$  &	$0.278$ &	$0.270$   &	$0.195$ \\ \midrule 
        	    \textbf{Q. Resolution~\cite{voskarides2020query}}       &	$0.391$ &	$0.300$ &	$0.315$  &	$0.232$ &	$0.213$   &	$0.220$ \\
        	    \textbf{Q. Rewriting~\cite{raposo2022question}}         &	$0.366$ &	$0.280$ &	$0.295$  &	$0.220$ &	$0.245$   &	$0.215$ \\ \midrule       	    
        	    \textbf{\convinse}                                      &	$\textbf{0.395}$ &	$\textbf{0.368}$* &	$0.320$  &	$0.298$ &	$\textbf{0.338}$*   &	$\textbf{0.252}$ \\
    		\bottomrule
    	\end{tabular} }
    \label{tab:per-turn}
    \vspace*{-0.3cm}
\end{table}

\myparagraph{\convinse is stable over turns} One striking finding when drilling down into the results, is that the performance of \convinse stays fairly stable as the conversation continues
-- see Table~\ref{tab:per-turn}. 
In contrast, as one would naturally expect, the baselines exhibit systematic degradation from turn to turn, as it becomes harder to capture implicit cues about the user intents with progressing conversation (this was the main reason for collecting \convmix-10T).
This is most pronounced for the \phrase{Prepend all} model.
Note, that different FiD models were trained for each method individually, which results in different performances even for the first turn.
For most methods, there is a significant performance drop for the last three turns, indicating that further investigation of generalization to longer conversations might be worthwhile.

\begin{table} [t] \small 
    \caption{Average QU output length in words on test sets.}
    \vspace*{-0.3cm}
	\begin{tabular}{l c c c}
	    \toprule
	    \textbf{QU Method}                  &   \textbf{\convmix-5T} & \textbf{\convmix-10T} & \textbf{\convmix}   \\ 
	    \toprule
	    \textbf{Original}                                   &   $5.73$  & $5.34$ & $5.57$ \\ 
	    \midrule
	    \textbf{Prepend init}                               &   $14.39$  & $14.84$ & $14.58$ \\ 
	    \textbf{Prepend prev}                               &   $12.23$  & $12.17$ & $12.21$ \\ 
	    \textbf{Prepend init+prev}                          &   $18.73$  & $20.61$ & $19.52$ \\ 
	    \textbf{Prepend all}                                &   $23.03$  & $40.70$ & $30.39$ \\  \midrule
	    \textbf{Q. Resolution~\cite{voskarides2020query}}   &   $8.24$  & $8.54$ & $8.36$  \\ 
	    \textbf{Q. Rewriting~\cite{raposo2022question}}     &   $9.07$  & $9.04$ & $9.06$   \\ 
	    \midrule
	    \textbf{\convinse}                                  &   $\textbf{6.43}$*  & $\textbf{6.53}$* & $\textbf{6.48}$* \\ 
	    \bottomrule
	\end{tabular}
	\label{tab:ques-len}
	\vspace*{-0.5cm}
\end{table}

\myparagraph{SRs are compact} 
SRs are indeed succinct representations of 
user intents, as indicated by Table~\ref{tab:ques-len} on question lengths. Also, for interpretability, SRs are an easy-to-inspect gist that are comprehensible to a human user and at the same time being amenable for use with most standard IR models.
Notably, for the more sophisticated models, the output length of the QU phase is almost stable on the two test sets with five turns and ten turns.

\myparagraph{\convinse is stable over domains} Zooming into the results over the five thematic domains in \convmix, we find that performance is relatively stable (working best for the movies domain) -- see Table~\ref{tab:per-domain} columns. The same holds when contrasting this with distributions over source types (Table~\ref{tab:per-domain} rows). Infoboxes consistently provide knowledge easily harnessed, while tables turn out to be the trickiest to handle, with proper verbalizations being a likely issue~\cite{oguz2021unikqa,thorne2020neural}.

\myparagraph{Slots in SR are vital} A systematic ablation study shows that each of the SR slots plays an important role -- see Table~\ref{tab:ablation}. We blanked out the contents of the respective slots during retrieval, and proceeded with these weaker SRs to the answering phase. Question entities clearly are the most pivotal; answer types do not help much at retrieval time, but justify their importance during answering (a shared insight w.r.t. expected answer types in many QA models~\cite{saharoy2021question}). We also examined the effect of our proposed ordering of the SR slots (e.g. predicates first).
As expected, there is hardly an effect during ERS (both \clocq and BM25 are word-order-agnostic), but ordering proves beneficial when generating answers from evidences using sequence-aware models like FiD.

\myparagraph{Error analysis}
\convinse cannot answer the question correctly $65.8$\% of the time, arising from three error cases:
i) the evidence retriever cannot retrieve any answering evidence ($42.4$\%), 
which can be due to the QU phase missing important cues in the conversation,
failures within the evidence retriever,
or the information sources not containing necessary information for answering,
ii) the evidence retriever retrieves at least one answering evidence
but none of these is among the top-$e$ after ERS ($27.2$\%), calling for more informed evidence-scoring mechanisms;
or iii) FiD fails to detect the correct answer ($30.4$\%), which indicates that more sophisticated reasoning over the set of evidences might be required.

\myparagraph{Anecdotal results} For a better understanding of how our SRs look like when contrasted with rewritten and resolved questions, we provide a few representative examples in Table~\ref{tab:anecdotes}.

\begin{table} \small
    \caption{Domain- and source-wise results (\convinse P@1).}
    \vspace*{-0.3cm}
    \newcolumntype{G}{>{\columncolor [gray] {0.90}}c}
    	\begin{tabular}{l c c c c c}
    		\toprule
    		    \textbf{Source}	 & \textbf{Books}  & \textbf{Movies} & \textbf{Music} & \textbf{TV Series} & \textbf{Soccer} \\
    		    \midrule
        	    \textbf{KB}	        &	$0.255$ &	$0.273$  &	$0.219$ &	$0.245$ &	$0.264$  \\
        	    \textbf{Text}	    &	$0.226$ &	$0.229$  &	$0.244$ &	$0.234$ &	$0.165$  \\
        	    \textbf{Table}	    &	$0.021$ &	$0.106$  &	$0.052$ &	$0.058$ &	$0.073$  \\
        	    \textbf{Info}	    &	$0.282$ &	$0.272$  &	$0.226$ &	$0.265$ &	$0.243$  \\
        	    \midrule
        	    \textbf{All}        &	$0.329$ &	$0.357$  &	$0.353$ &	$0.338$ &	$0.333$  \\
    		\bottomrule
    	\end{tabular} 
    \label{tab:per-domain}
\end{table}

\begin{table} \small
    \caption{Ablation study of the \convinse SR.}
    \vspace*{-0.3cm}
    \newcolumntype{G}{>{\columncolor [gray] {0.90}}c}
    	\begin{tabular}{l c c c c G}
    		\toprule
                \textbf{Method}                             & \textbf{Answer presence (ERS)}	 & \textbf{P@1 (HA)} \\ \toprule
        	    \textbf{\convinse}                          & $0.559$                 & $\textbf{0.371}$    \\ \midrule
        	    \textbf{w/o Context entity slot}	        & $0.546$                 & $0.362$                 \\
        	    \textbf{w/o Question entity slot}	        & $0.078$                 & $0.054$                 \\
        	    \textbf{w/o Predicate slot}	                & $0.421$                 & $0.176$                 \\        	    
        	    \textbf{w/o Answer type slot}           	& $\textbf{0.572}$    & $0.350$                 \\
        	    \textbf{w/o Ordering}	                    & $0.560$                 &	$0.361$                 \\        	    
    		\bottomrule
    	\end{tabular} 
    \label{tab:ablation}
\end{table}

\begin{table} [t] \small 
    \caption{Cases where only \convinse answered correctly.}
    \vspace*{-0.3cm}
	\begin{tabular}{p{1cm} p{6.7cm}}
	    \toprule
	    \textbf{Domain} & \textbf{Books} \\ \midrule
	    \textbf{Original} & country of origin? \\
	    \textbf{Q. Res.}    & country of origin? expanse leviathan  \\
	    \textbf{Q. Rew.}     & What country was it taken from?     \\
	    \textbf{SR}           & $\langle$ \struct{\textcolor{gray}{Leviathan Wakes} | \textcolor{red}{Expanse} | \textcolor{blue}{country origin} | \textcolor{cyan}{sovereign state}} $\rangle$ \\ \midrule
        \textbf{Domain} & \textbf{Movies} \\ \midrule
        \textbf{Original} & What actor portrayed Magneto? \\
        \textbf{Q. Res.}    &  What actor portrayed Magneto? x-men movie   \\
	    \textbf{Q. Rew.}     & Which actor played the character Magneto in the X-Men movie?   \\
	    \textbf{SR}           & $\langle$ \struct{\textcolor{gray}{X-Men} | \textcolor{red}{Magneto} | \textcolor{blue}{actor portrayed} | \textcolor{cyan}{human}} $\rangle$  \\ \midrule
	    \textbf{Domain} & \textbf{Music} \\ \midrule
	    \textbf{Original} & Her date of birth? \\
	    \textbf{Q. Res.}    & Her date of birth? shakira                 \\
	    \textbf{Q. Rew.}     & When was Shakira born?     \\
	    \textbf{SR}           & $\langle$ \struct{\textcolor{gray}{Waka Waka (This Time for Africa)} | \textcolor{red}{Shakira}} \\
	    & \struct{ |  \textcolor{blue}{date birth} | \textcolor{cyan}{date}}  $\rangle$  \\ \midrule
	    \textbf{Domain} & \textbf{TV Series} \\ \midrule
	    \textbf{Original} & What TV series did he first appear on? \\
	    \textbf{Q. Res.}    &  What TV series did he first appear on? tv show appeared the shows george   \\
	    \textbf{Q. Rew.}     &  What TV series did George Grizzard first appear on?   \\
	    \textbf{SR}           & $\langle$ \struct{\textcolor{gray}{\_} | \textcolor{red}{George Grizzard} | \textcolor{blue}{tv series first appear} | \textcolor{cyan}{television series}} $\rangle$  \\\midrule
	    \textbf{Domain} & \textbf{Soccer} \\ \midrule
	    	    \textbf{Original} & Who won? \\
	    \textbf{Q. Res.}    & Who won? shakira                 \\
	    \textbf{Q. Rew.}     & Who won the FIFA FIFA World Cup in France?     \\
	    \textbf{SR}           & $\langle$ \struct{\textcolor{gray}{\_} | \textcolor{red}{2010 FIFA World Cup} |  \textcolor{blue}{won} | \textcolor{cyan}{national association football team}} $\rangle$  \\
	    \bottomrule
	\end{tabular}
	\label{tab:anecdotes}
    \vspace*{-0.4cm}
\end{table}

%% file: sections/07-related.tex
\section{Related Work}
\label{sec:related}

\myparagraph{Conversational question answering} Some methods for ConvQA over text use a \textit{hard history selection} to obtain a question-relevant subset of the conversation history~\cite{qiu2021reinforced, qu2019bert, kaiser2020conversational}.
However, the more prevalent
approach is to construct a question-aware encoding
with attention
on the conversational history (\textit{soft history selection})~\cite{chengraphflow, huang2018flowqa, qu2019attentive}.
This is then used by a 
neural reader
for predicting the answer in the given passage. 
While this works well in ConvQA over text, 
the whole pipeline does not easily generalize to the heterogeneous setting,
since this would require carefully designed mechanisms for representing inputs from different sources in a shared latent space.
%
It may seem that hard history selection can be generalized to the heterogeneous setting more easily, but we found that such an approach hurts
end-to-end answering performance in preliminary experiments.
We employ attention over conversation history in the QU phase, which can be seen as a soft history selection.

In ConvQA over KBs, the conversational flow is typically captured in a context graph, either explicity~\cite{christmann2019look} using iterative expansion, or implicitly~\cite{kaiser2021reinforcement, guo2018dialog, marion2021structured} by maintaining a set of context entities and predicates.
Conversational KB history can be incorporated using encodings~\cite{guo2018dialog, shen2019multi, marion2021structured} or heuristic subsets of the history~\cite{kaiser2021reinforcement}, similar to ConvQA over text.
All existing methods consider only one information source for answering conversations, which limits their answer coverage.
Moreover, 
such works
often adopt source-specific modeling of the conversational flow, like KB subgraphs, and
cannot easily be extended to capture the conversational flow in a heterogeneous setting.

\myparagraph{Question completion}
One line of work that suggests a more general approach to the ConvQA problem 
aims to create
a
self-contained question
from the incomplete utterance
that can be answered by standalone QA systems~\cite{vakulenko2021question, xu2020learning, elgohary2019can, voskarides2020query, raposo2022question}.
Such approaches can either take the form of 
question rewriting,
which generates a complete question from scratch~\cite{vakulenko2021question, raposo2022question},
or question resolution, which adds
relevant terms from the conversational history to the question~\cite{voskarides2020query}.
Question rewriting can entail unnecessary complexity
since recent QA systems may not require the question to have perfect syntax or grammar~\cite{izacard2021leveraging,jia21complex}.

Question resolution~\cite{voskarides2020query} operates on the surface level, without aiming for grammatically correct questions, creating a completed question similar to a keyword query.
However, a potential downside is that there is no structure in the completed form.

We show in experiments (Sec.~\ref{subsec:main-res})
that both, completed forms via question rewriting, and question resolution,
perform worse than our generated
structured representations,
which can be perceived as logical forms for matching with heterogeneous sources.

\myparagraph{Heterogeneous answering}
There has been work on combining text and KBs~\cite{xiong2019improving,sun2018open,sun2019pullnet,savenkov2016knowledge,xu2016question,pramanik2021uniqorn} for conventional QA with single self-contained questions.
O\u{g}uz et al.~\cite{oguz2021unikqa} and Ma et al.~\cite{ma2021open} integrate tables as
an additional type of information source.
There has also been work on combining text and tables~\cite{chen2020hybridqa, chen2020open, zhu2021tat}, or text, tables and images~\cite{talmor2021multimodalqa, hannan2020manymodalqa} for answering full-fledged questions.
However, these approaches cannot easily be extended to a conversational setting with incomplete follow-up questions.

%% file: sections/08-conclusion.tex

\section{Conclusions and future work}
\label{sec:confut}

We presented \convinse, the first end-to-end framework for answering conversational questions over heterogeneous sources, covering knowledge bases (KBs), text corpora, Web tables, and Wikipedia infoboxes in a seamless manner. 
Our approach marries ideas from different communities and
their methodological paradigms. The learning of intent-explicit
structured representations (SRs) is inspired by semantic frames in symbolic AI,
but leaves its constituents in surface form without normalization.
For retrieving relevant evidences from heterogeneous sources,
we follow the line of IR thinking. The 
design of SRs is vital here: the SRs provide enough structure to
enhance effectiveness, but are lightweight and can
be directly applied to KBs, text, tables, and infoboxes alike.
Finally, the last stage of extracting answers adopts the NLP paradigm of QA over text.
%
%
We believe that 
extending conversational QA into a heterogeneous retrieval problem is a useful contribution to the community.
The proposed \convmix benchmark, with thousands of real-user conversations, will
help to further advance
the state-of-the-art.
One of the most promising avenues for future work involves developing more informed retrieval and answering phases, making direct use of the semantic roles in structured representations.

\vspace*{0.3cm}
\myparagraph{Acknowledgments}
We
thank 
Magdalena Kaiser from the MPI for Informatics for her useful inputs on this work. 
Part of this work was supported by the
ERC Synergy Grant ``imPACT'' (No. 610150).

%% file: 2022-sigir-fp-convqa.bbl

\begin{thebibliography}{59}


\ifx \showCODEN    \undefined \def \showCODEN     #1{\unskip}     \fi
\ifx \showDOI      \undefined \def \showDOI       #1{#1}\fi
\ifx \showISBNx    \undefined \def \showISBNx     #1{\unskip}     \fi
\ifx \showISBNxiii \undefined \def \showISBNxiii  #1{\unskip}     \fi
\ifx \showISSN     \undefined \def \showISSN      #1{\unskip}     \fi
\ifx \showLCCN     \undefined \def \showLCCN      #1{\unskip}     \fi
\ifx \shownote     \undefined \def \shownote      #1{#1}          \fi
\ifx \showarticletitle \undefined \def \showarticletitle #1{#1}   \fi
\ifx \showURL      \undefined \def \showURL       {\relax}        \fi
\providecommand\bibfield[2]{#2}
\providecommand\bibinfo[2]{#2}
\providecommand\natexlab[1]{#1}
\providecommand\showeprint[2][]{arXiv:#2}

\bibitem[\protect\citeauthoryear{Auer, Bizer, Kobilarov, Lehmann, Cyganiak, and
  Ives}{Auer et~al\mbox{.}}{2007}]%
        {auer2007dbpedia}
\bibfield{author}{\bibinfo{person}{S{\"o}ren Auer}, \bibinfo{person}{Christian
  Bizer}, \bibinfo{person}{Georgi Kobilarov}, \bibinfo{person}{Jens Lehmann},
  \bibinfo{person}{Richard Cyganiak}, {and} \bibinfo{person}{Zachary Ives}.}
  \bibinfo{year}{2007}\natexlab{}.
\newblock \showarticletitle{{DBpedia: A nucleus for a Web of open data}}. In
  \bibinfo{booktitle}{\emph{Sem. Web}}.
\newblock


\bibitem[\protect\citeauthoryear{Chen, Fisch, Weston, and Bordes}{Chen
  et~al\mbox{.}}{2017}]%
        {chen2017reading}
\bibfield{author}{\bibinfo{person}{Danqi Chen}, \bibinfo{person}{Adam Fisch},
  \bibinfo{person}{Jason Weston}, {and} \bibinfo{person}{Antoine Bordes}.}
  \bibinfo{year}{2017}\natexlab{}.
\newblock \showarticletitle{Reading Wikipedia to Answer Open-Domain Questions}.
  In \bibinfo{booktitle}{\emph{ACL}}.
\newblock


\bibitem[\protect\citeauthoryear{Chen, Chang, Schlinger, Wang, and Cohen}{Chen
  et~al\mbox{.}}{2021}]%
        {chen2020open}
\bibfield{author}{\bibinfo{person}{Wenhu Chen}, \bibinfo{person}{Ming-Wei
  Chang}, \bibinfo{person}{Eva Schlinger}, \bibinfo{person}{William Wang},
  {and} \bibinfo{person}{William~W Cohen}.} \bibinfo{year}{2021}\natexlab{}.
\newblock \showarticletitle{Open question answering over tables and text}. In
  \bibinfo{booktitle}{\emph{ICLR}}.
\newblock


\bibitem[\protect\citeauthoryear{Chen, Zha, Chen, Xiong, Wang, and Wang}{Chen
  et~al\mbox{.}}{2020b}]%
        {chen2020hybridqa}
\bibfield{author}{\bibinfo{person}{Wenhu Chen}, \bibinfo{person}{Hanwen Zha},
  \bibinfo{person}{Zhiyu Chen}, \bibinfo{person}{Wenhan Xiong},
  \bibinfo{person}{Hong Wang}, {and} \bibinfo{person}{William~Yang Wang}.}
  \bibinfo{year}{2020}\natexlab{b}.
\newblock \showarticletitle{HybridQA: A Dataset of Multi-Hop Question Answering
  over Tabular and Textual Data}. In \bibinfo{booktitle}{\emph{EMNLP}}.
\newblock


\bibitem[\protect\citeauthoryear{Chen, Wu, and Zaki}{Chen
  et~al\mbox{.}}{2020a}]%
        {chengraphflow}
\bibfield{author}{\bibinfo{person}{Yu Chen}, \bibinfo{person}{Lingfei Wu},
  {and} \bibinfo{person}{Mohammed~J Zaki}.} \bibinfo{year}{2020}\natexlab{a}.
\newblock \showarticletitle{GraphFlow: Exploiting Conversation Flow with Graph
  Neural Networks for Conversational Machine Comprehension}. In
  \bibinfo{booktitle}{\emph{IJCAI}}.
\newblock


\bibitem[\protect\citeauthoryear{Choi, He, Iyyer, Yatskar, Yih, Choi, Liang,
  and Zettlemoyer}{Choi et~al\mbox{.}}{2018}]%
        {choi2018quac}
\bibfield{author}{\bibinfo{person}{Eunsol Choi}, \bibinfo{person}{He He},
  \bibinfo{person}{Mohit Iyyer}, \bibinfo{person}{Mark Yatskar},
  \bibinfo{person}{Wen-tau Yih}, \bibinfo{person}{Yejin Choi},
  \bibinfo{person}{Percy Liang}, {and} \bibinfo{person}{Luke Zettlemoyer}.}
  \bibinfo{year}{2018}\natexlab{}.
\newblock \showarticletitle{{QuAC: Q}uestion answering in context}. In
  \bibinfo{booktitle}{\emph{EMNLP}}.
\newblock


\bibitem[\protect\citeauthoryear{Christmann, Roy, and Weikum}{Christmann
  et~al\mbox{.}}{2022}]%
        {christmann2022beyond}
\bibfield{author}{\bibinfo{person}{Philipp Christmann},
  \bibinfo{person}{Rishiraj~Saha Roy}, {and} \bibinfo{person}{Gerhard Weikum}.}
  \bibinfo{year}{2022}\natexlab{}.
\newblock \showarticletitle{{Beyond NED: F}ast and Effective Search Space
  Reduction for Complex Question Answering over Knowledge Bases}. In
  \bibinfo{booktitle}{\emph{WSDM}}.
\newblock


\bibitem[\protect\citeauthoryear{Christmann, Saha~Roy, Abujabal, Singh, and
  Weikum}{Christmann et~al\mbox{.}}{2019}]%
        {christmann2019look}
\bibfield{author}{\bibinfo{person}{Philipp Christmann},
  \bibinfo{person}{Rishiraj Saha~Roy}, \bibinfo{person}{Abdalghani Abujabal},
  \bibinfo{person}{Jyotsna Singh}, {and} \bibinfo{person}{Gerhard Weikum}.}
  \bibinfo{year}{2019}\natexlab{}.
\newblock \showarticletitle{Look before you hop: Conversational question
  answering over knowledge graphs using judicious context expansion}. In
  \bibinfo{booktitle}{\emph{CIKM}}.
\newblock


\bibitem[\protect\citeauthoryear{Elgohary, Peskov, and Boyd-Graber}{Elgohary
  et~al\mbox{.}}{2019}]%
        {elgohary2019can}
\bibfield{author}{\bibinfo{person}{Ahmed Elgohary}, \bibinfo{person}{Denis
  Peskov}, {and} \bibinfo{person}{Jordan Boyd-Graber}.}
  \bibinfo{year}{2019}\natexlab{}.
\newblock \showarticletitle{Can You Unpack That? Learning to Rewrite
  Questions-in-Context}. In \bibinfo{booktitle}{\emph{EMNLP-IJCNLP}}.
\newblock


\bibitem[\protect\citeauthoryear{Guo, Tang, Duan, Zhou, and Yin}{Guo
  et~al\mbox{.}}{2018}]%
        {guo2018dialog}
\bibfield{author}{\bibinfo{person}{Daya Guo}, \bibinfo{person}{Duyu Tang},
  \bibinfo{person}{Nan Duan}, \bibinfo{person}{Ming Zhou}, {and}
  \bibinfo{person}{Jian Yin}.} \bibinfo{year}{2018}\natexlab{}.
\newblock \showarticletitle{Dialog-to-action: conversational question answering
  over a large-scale knowledge base}. In \bibinfo{booktitle}{\emph{NeurIPS}}.
\newblock


\bibitem[\protect\citeauthoryear{Gupta and Bendersky}{Gupta and
  Bendersky}{2015}]%
        {gupta2015information}
\bibfield{author}{\bibinfo{person}{Manish Gupta} {and} \bibinfo{person}{Michael
  Bendersky}.} \bibinfo{year}{2015}\natexlab{}.
\newblock \showarticletitle{Information retrieval with verbose queries}. In
  \bibinfo{booktitle}{\emph{SIGIR}}.
\newblock


\bibitem[\protect\citeauthoryear{Hannan, Jain, and Bansal}{Hannan
  et~al\mbox{.}}{2020}]%
        {hannan2020manymodalqa}
\bibfield{author}{\bibinfo{person}{Darryl Hannan}, \bibinfo{person}{Akshay
  Jain}, {and} \bibinfo{person}{Mohit Bansal}.}
  \bibinfo{year}{2020}\natexlab{}.
\newblock \showarticletitle{ManyModalQA: Modality Disambiguation and QA over
  Diverse Inputs}. In \bibinfo{booktitle}{\emph{AAAI}},
  Vol.~\bibinfo{volume}{34}.
\newblock


\bibitem[\protect\citeauthoryear{Ho, Ibrahim, Pal, Berberich, and Weikum}{Ho
  et~al\mbox{.}}{2019}]%
        {ho2019qsearch}
\bibfield{author}{\bibinfo{person}{Vinh~Thinh Ho}, \bibinfo{person}{Yusra
  Ibrahim}, \bibinfo{person}{Koninika Pal}, \bibinfo{person}{Klaus Berberich},
  {and} \bibinfo{person}{Gerhard Weikum}.} \bibinfo{year}{2019}\natexlab{}.
\newblock \showarticletitle{Qsearch: Answering quantity queries from text}. In
  \bibinfo{booktitle}{\emph{ISWC}}.
\newblock


\bibitem[\protect\citeauthoryear{Huang, Choi, and Yih}{Huang
  et~al\mbox{.}}{2018}]%
        {huang2018flowqa}
\bibfield{author}{\bibinfo{person}{Hsin-Yuan Huang}, \bibinfo{person}{Eunsol
  Choi}, {and} \bibinfo{person}{Wen-tau Yih}.} \bibinfo{year}{2018}\natexlab{}.
\newblock \showarticletitle{FlowQA: Grasping Flow in History for Conversational
  Machine Comprehension}. In \bibinfo{booktitle}{\emph{ICLR}}.
\newblock


\bibitem[\protect\citeauthoryear{Iyyer, Yih, and Chang}{Iyyer
  et~al\mbox{.}}{2017}]%
        {iyyer2017search}
\bibfield{author}{\bibinfo{person}{Mohit Iyyer}, \bibinfo{person}{Wen-tau Yih},
  {and} \bibinfo{person}{Ming-Wei Chang}.} \bibinfo{year}{2017}\natexlab{}.
\newblock \showarticletitle{Search-based neural structured learning for
  sequential question answering}. In \bibinfo{booktitle}{\emph{ACL}}.
\newblock


\bibitem[\protect\citeauthoryear{Izacard and Grave}{Izacard and Grave}{2021}]%
        {izacard2021leveraging}
\bibfield{author}{\bibinfo{person}{Gautier Izacard} {and}
  \bibinfo{person}{{\'E}douard Grave}.} \bibinfo{year}{2021}\natexlab{}.
\newblock \showarticletitle{Leveraging Passage Retrieval with Generative Models
  for Open Domain Question Answering}. In \bibinfo{booktitle}{\emph{EACL}}.
\newblock


\bibitem[\protect\citeauthoryear{Jia, Pramanik, Saha~Roy, and Weikum}{Jia
  et~al\mbox{.}}{2021}]%
        {jia21complex}
\bibfield{author}{\bibinfo{person}{Zhen Jia}, \bibinfo{person}{Soumajit
  Pramanik}, \bibinfo{person}{Rishiraj Saha~Roy}, {and}
  \bibinfo{person}{Gerhard Weikum}.} \bibinfo{year}{2021}\natexlab{}.
\newblock \showarticletitle{Complex Temporal Question Answering on Knowledge
  Graphs}. In \bibinfo{booktitle}{\emph{CIKM}}.
\newblock


\bibitem[\protect\citeauthoryear{Joko, Hasibi, Balog, and de~Vries}{Joko
  et~al\mbox{.}}{2021}]%
        {joko2021conversational}
\bibfield{author}{\bibinfo{person}{Hideaki Joko}, \bibinfo{person}{Faegheh
  Hasibi}, \bibinfo{person}{Krisztian Balog}, {and} \bibinfo{person}{Arjen~P de
  Vries}.} \bibinfo{year}{2021}\natexlab{}.
\newblock \showarticletitle{Conversational Entity Linking: Problem Definition
  and Datasets}. In \bibinfo{booktitle}{\emph{SIGIR}}.
\newblock


\bibitem[\protect\citeauthoryear{Kacupaj, Plepi, Singh, Thakkar, Lehmann, and
  Maleshkova}{Kacupaj et~al\mbox{.}}{2021}]%
        {kacupaj2021conversational}
\bibfield{author}{\bibinfo{person}{Endri Kacupaj}, \bibinfo{person}{Joan
  Plepi}, \bibinfo{person}{Kuldeep Singh}, \bibinfo{person}{Harsh Thakkar},
  \bibinfo{person}{Jens Lehmann}, {and} \bibinfo{person}{Maria Maleshkova}.}
  \bibinfo{year}{2021}\natexlab{}.
\newblock \showarticletitle{Conversational Question Answering over Knowledge
  Graphs with Transformer and Graph Attention Networks}. In
  \bibinfo{booktitle}{\emph{EACL}}.
\newblock


\bibitem[\protect\citeauthoryear{Kaiser, Saha~Roy, and Weikum}{Kaiser
  et~al\mbox{.}}{2020}]%
        {kaiser2020conversational}
\bibfield{author}{\bibinfo{person}{Magdalena Kaiser}, \bibinfo{person}{Rishiraj
  Saha~Roy}, {and} \bibinfo{person}{Gerhard Weikum}.}
  \bibinfo{year}{2020}\natexlab{}.
\newblock \showarticletitle{Conversational Question Answering over Passages by
  Leveraging Word Proximity Networks}. In \bibinfo{booktitle}{\emph{SIGIR}}.
\newblock


\bibitem[\protect\citeauthoryear{Kaiser, Saha~Roy, and Weikum}{Kaiser
  et~al\mbox{.}}{2021}]%
        {kaiser2021reinforcement}
\bibfield{author}{\bibinfo{person}{Magdalena Kaiser}, \bibinfo{person}{Rishiraj
  Saha~Roy}, {and} \bibinfo{person}{Gerhard Weikum}.}
  \bibinfo{year}{2021}\natexlab{}.
\newblock \showarticletitle{Reinforcement Learning from Reformulations in
  Conversational Question Answering over Knowledge Graphs}. In
  \bibinfo{booktitle}{\emph{SIGIR}}.
\newblock


\bibitem[\protect\citeauthoryear{Kumar and Joshi}{Kumar and Joshi}{2017}]%
        {kumar2017incomplete}
\bibfield{author}{\bibinfo{person}{Vineet Kumar} {and}
  \bibinfo{person}{Sachindra Joshi}.} \bibinfo{year}{2017}\natexlab{}.
\newblock \showarticletitle{Incomplete follow-up question resolution using
  retrieval based sequence to sequence learning}. In
  \bibinfo{booktitle}{\emph{SIGIR}}.
\newblock


\bibitem[\protect\citeauthoryear{Lan and Jiang}{Lan and Jiang}{2021}]%
        {lan2021modeling}
\bibfield{author}{\bibinfo{person}{Yunshi Lan} {and} \bibinfo{person}{Jing
  Jiang}.} \bibinfo{year}{2021}\natexlab{}.
\newblock \showarticletitle{Modeling Transitions of Focal Entities for
  Conversational Knowledge Base Question Answering}. In
  \bibinfo{booktitle}{\emph{ACL}}.
\newblock


\bibitem[\protect\citeauthoryear{Levenshtein et~al\mbox{.}}{Levenshtein
  et~al\mbox{.}}{1966}]%
        {levenshtein1966binary}
\bibfield{author}{\bibinfo{person}{Vladimir~I Levenshtein} {et~al\mbox{.}}}
  \bibinfo{year}{1966}\natexlab{}.
\newblock \showarticletitle{Binary codes capable of correcting deletions,
  insertions, and reversals}. In \bibinfo{booktitle}{\emph{Soviet physics
  doklady}}.
\newblock


\bibitem[\protect\citeauthoryear{Lewis, Liu, Goyal, Ghazvininejad, Mohamed,
  Levy, Stoyanov, and Zettlemoyer}{Lewis et~al\mbox{.}}{2020}]%
        {lewis2020bart}
\bibfield{author}{\bibinfo{person}{Mike Lewis}, \bibinfo{person}{Yinhan Liu},
  \bibinfo{person}{Naman Goyal}, \bibinfo{person}{Marjan Ghazvininejad},
  \bibinfo{person}{Abdelrahman Mohamed}, \bibinfo{person}{Omer Levy},
  \bibinfo{person}{Veselin Stoyanov}, {and} \bibinfo{person}{Luke
  Zettlemoyer}.} \bibinfo{year}{2020}\natexlab{}.
\newblock \showarticletitle{BART: Denoising Sequence-to-Sequence Pre-training
  for Natural Language Generation, Translation, and Comprehension}. In
  \bibinfo{booktitle}{\emph{ACL}}.
\newblock


\bibitem[\protect\citeauthoryear{Ma, Cheng, Liu, Nyberg, and Gao}{Ma
  et~al\mbox{.}}{2022}]%
        {ma2021open}
\bibfield{author}{\bibinfo{person}{Kaixin Ma}, \bibinfo{person}{Hao Cheng},
  \bibinfo{person}{Xiaodong Liu}, \bibinfo{person}{Eric Nyberg}, {and}
  \bibinfo{person}{Jianfeng Gao}.} \bibinfo{year}{2022}\natexlab{}.
\newblock \showarticletitle{{Open Domain Question Answering with A Unified
  Knowledge Interface}}. In \bibinfo{booktitle}{\emph{ACL}}.
\newblock


\bibitem[\protect\citeauthoryear{Marion, Nowak, and Piccinno}{Marion
  et~al\mbox{.}}{2021}]%
        {marion2021structured}
\bibfield{author}{\bibinfo{person}{Pierre Marion},
  \bibinfo{person}{Pawe{\l}~Krzysztof Nowak}, {and} \bibinfo{person}{Francesco
  Piccinno}.} \bibinfo{year}{2021}\natexlab{}.
\newblock \showarticletitle{Structured Context and High-Coverage Grammar for
  Conversational Question Answering over Knowledge Graphs}. In
  \bibinfo{booktitle}{\emph{EMNLP}}.
\newblock


\bibitem[\protect\citeauthoryear{Mueller, Piccinno, Shaw, Nicosia, and
  Altun}{Mueller et~al\mbox{.}}{2019}]%
        {mueller2019answering}
\bibfield{author}{\bibinfo{person}{Thomas Mueller}, \bibinfo{person}{Francesco
  Piccinno}, \bibinfo{person}{Peter Shaw}, \bibinfo{person}{Massimo Nicosia},
  {and} \bibinfo{person}{Yasemin Altun}.} \bibinfo{year}{2019}\natexlab{}.
\newblock \showarticletitle{Answering Conversational Questions on Structured
  Data without Logical Forms}. In \bibinfo{booktitle}{\emph{EMNLP-IJCNLP}}.
\newblock


\bibitem[\protect\citeauthoryear{O\u{g}uz, Chen, Karpukhin, Peshterliev,
  Okhonko, Schlichtkrull, Gupta, Mehdad, and Yih}{O\u{g}uz
  et~al\mbox{.}}{2021}]%
        {oguz2021unikqa}
\bibfield{author}{\bibinfo{person}{Barlas O\u{g}uz}, \bibinfo{person}{Xilun
  Chen}, \bibinfo{person}{Vladimir Karpukhin}, \bibinfo{person}{Stan
  Peshterliev}, \bibinfo{person}{Dmytro Okhonko}, \bibinfo{person}{Michael
  Schlichtkrull}, \bibinfo{person}{Sonal Gupta}, \bibinfo{person}{Yashar
  Mehdad}, {and} \bibinfo{person}{Scott Yih}.} \bibinfo{year}{2021}\natexlab{}.
\newblock \showarticletitle{{UniK-QA: Unified Representations of Structured and
  Unstructured Knowledge for Open-Domain Question Answering}}.
\newblock \bibinfo{journal}{\emph{arXiv}} (\bibinfo{year}{2021}).
\newblock


\bibitem[\protect\citeauthoryear{Plepi, Kacupaj, Singh, Thakkar, and
  Lehmann}{Plepi et~al\mbox{.}}{2021}]%
        {plepi2021context}
\bibfield{author}{\bibinfo{person}{Joan Plepi}, \bibinfo{person}{Endri
  Kacupaj}, \bibinfo{person}{Kuldeep Singh}, \bibinfo{person}{Harsh Thakkar},
  {and} \bibinfo{person}{Jens Lehmann}.} \bibinfo{year}{2021}\natexlab{}.
\newblock \showarticletitle{Context Transformer with Stacked Pointer Networks
  for Conversational Question Answering over Knowledge Graphs}. In
  \bibinfo{booktitle}{\emph{ESWC}}.
\newblock


\bibitem[\protect\citeauthoryear{Pramanik, Alabi, Roy, and Weikum}{Pramanik
  et~al\mbox{.}}{2021}]%
        {pramanik2021uniqorn}
\bibfield{author}{\bibinfo{person}{Soumajit Pramanik},
  \bibinfo{person}{Jesujoba Alabi}, \bibinfo{person}{Rishiraj~Saha Roy}, {and}
  \bibinfo{person}{Gerhard Weikum}.} \bibinfo{year}{2021}\natexlab{}.
\newblock \showarticletitle{{UNIQORN: Unified Question Answering over RDF
  Knowledge Graphs and Natural Language Text}}.
\newblock \bibinfo{journal}{\emph{arXiv}} (\bibinfo{year}{2021}).
\newblock


\bibitem[\protect\citeauthoryear{Qiu, Huang, Chen, Ji, Qu, Wei, Huang, and
  Zhang}{Qiu et~al\mbox{.}}{2021}]%
        {qiu2021reinforced}
\bibfield{author}{\bibinfo{person}{Minghui Qiu}, \bibinfo{person}{Xinjing
  Huang}, \bibinfo{person}{Cen Chen}, \bibinfo{person}{Feng Ji},
  \bibinfo{person}{Chen Qu}, \bibinfo{person}{Wei Wei}, \bibinfo{person}{Jun
  Huang}, {and} \bibinfo{person}{Yin Zhang}.} \bibinfo{year}{2021}\natexlab{}.
\newblock \showarticletitle{Reinforced History Backtracking for Conversational
  Question Answering}. In \bibinfo{booktitle}{\emph{AAAI}},
  Vol.~\bibinfo{volume}{35}.
\newblock


\bibitem[\protect\citeauthoryear{Qu, Yang, Chen, Qiu, Croft, and Iyyer}{Qu
  et~al\mbox{.}}{2020}]%
        {qu2020open}
\bibfield{author}{\bibinfo{person}{Chen Qu}, \bibinfo{person}{Liu Yang},
  \bibinfo{person}{Cen Chen}, \bibinfo{person}{Minghui Qiu},
  \bibinfo{person}{W~Bruce Croft}, {and} \bibinfo{person}{Mohit Iyyer}.}
  \bibinfo{year}{2020}\natexlab{}.
\newblock \showarticletitle{Open-retrieval conversational question answering}.
  In \bibinfo{booktitle}{\emph{SIGIR}}.
\newblock


\bibitem[\protect\citeauthoryear{Qu, Yang, Qiu, Croft, Zhang, and Iyyer}{Qu
  et~al\mbox{.}}{2019a}]%
        {qu2019bert}
\bibfield{author}{\bibinfo{person}{Chen Qu}, \bibinfo{person}{Liu Yang},
  \bibinfo{person}{Minghui Qiu}, \bibinfo{person}{W~Bruce Croft},
  \bibinfo{person}{Yongfeng Zhang}, {and} \bibinfo{person}{Mohit Iyyer}.}
  \bibinfo{year}{2019}\natexlab{a}.
\newblock \showarticletitle{BERT with history answer embedding for
  conversational question answering}. In \bibinfo{booktitle}{\emph{SIGIR}}.
\newblock


\bibitem[\protect\citeauthoryear{Qu, Yang, Qiu, Zhang, Chen, Croft, and
  Iyyer}{Qu et~al\mbox{.}}{2019b}]%
        {qu2019attentive}
\bibfield{author}{\bibinfo{person}{Chen Qu}, \bibinfo{person}{Liu Yang},
  \bibinfo{person}{Minghui Qiu}, \bibinfo{person}{Yongfeng Zhang},
  \bibinfo{person}{Cen Chen}, \bibinfo{person}{W~Bruce Croft}, {and}
  \bibinfo{person}{Mohit Iyyer}.} \bibinfo{year}{2019}\natexlab{b}.
\newblock \showarticletitle{Attentive history selection for conversational
  question answering}. In \bibinfo{booktitle}{\emph{CIKM}}.
\newblock


\bibitem[\protect\citeauthoryear{Raffel, Shazeer, Roberts, Lee, Narang, Matena,
  Zhou, Li, and Liu}{Raffel et~al\mbox{.}}{2020}]%
        {raffel2020exploring}
\bibfield{author}{\bibinfo{person}{Colin Raffel}, \bibinfo{person}{Noam
  Shazeer}, \bibinfo{person}{Adam Roberts}, \bibinfo{person}{Katherine Lee},
  \bibinfo{person}{Sharan Narang}, \bibinfo{person}{Michael Matena},
  \bibinfo{person}{Yanqi Zhou}, \bibinfo{person}{Wei Li}, {and}
  \bibinfo{person}{Peter~J Liu}.} \bibinfo{year}{2020}\natexlab{}.
\newblock \showarticletitle{Exploring the limits of transfer learning with a
  unified text-to-text transformer}.
\newblock \bibinfo{journal}{\emph{JMLR}} (\bibinfo{year}{2020}).
\newblock


\bibitem[\protect\citeauthoryear{Raposo, Ribeiro, Martins, and Coheur}{Raposo
  et~al\mbox{.}}{2022}]%
        {raposo2022question}
\bibfield{author}{\bibinfo{person}{Gon\c{c}alo Raposo}, \bibinfo{person}{Rui
  Ribeiro}, \bibinfo{person}{Bruno Martins}, {and} \bibinfo{person}{Luísa
  Coheur}.} \bibinfo{year}{2022}\natexlab{}.
\newblock \showarticletitle{{Question rewriting? A}ssessing its importance for
  conversational question answering}. In \bibinfo{booktitle}{\emph{ECIR}}.
\newblock


\bibitem[\protect\citeauthoryear{Reddy, Chen, and Manning}{Reddy
  et~al\mbox{.}}{2019}]%
        {reddy2019coqa}
\bibfield{author}{\bibinfo{person}{Siva Reddy}, \bibinfo{person}{Danqi Chen},
  {and} \bibinfo{person}{Christopher Manning}.}
  \bibinfo{year}{2019}\natexlab{}.
\newblock \showarticletitle{{CoQA: A} conversational question answering
  challenge}.
\newblock \bibinfo{journal}{\emph{TACL}} (\bibinfo{year}{2019}).
\newblock


\bibitem[\protect\citeauthoryear{Robertson and Zaragoza}{Robertson and
  Zaragoza}{2009}]%
        {robertson2009probabilistic}
\bibfield{author}{\bibinfo{person}{Stephen Robertson} {and}
  \bibinfo{person}{Hugo Zaragoza}.} \bibinfo{year}{2009}\natexlab{}.
\newblock \showarticletitle{The Probabilistic Relevance Framework: BM25 and
  Beyond}.
\newblock \bibinfo{journal}{\emph{Foundations and Trends in Information
  Retrieval}} (\bibinfo{year}{2009}).
\newblock


\bibitem[\protect\citeauthoryear{Roller, Dinan, Goyal, Ju, Williamson, Liu, Xu,
  Ott, Smith, Boureau, et~al\mbox{.}}{Roller et~al\mbox{.}}{2021}]%
        {roller2021recipes}
\bibfield{author}{\bibinfo{person}{Stephen Roller}, \bibinfo{person}{Emily
  Dinan}, \bibinfo{person}{Naman Goyal}, \bibinfo{person}{Da Ju},
  \bibinfo{person}{Mary Williamson}, \bibinfo{person}{Yinhan Liu},
  \bibinfo{person}{Jing Xu}, \bibinfo{person}{Myle Ott},
  \bibinfo{person}{Eric~Michael Smith}, \bibinfo{person}{Y-Lan Boureau},
  {et~al\mbox{.}}} \bibinfo{year}{2021}\natexlab{}.
\newblock \showarticletitle{Recipes for Building an Open-Domain Chatbot}. In
  \bibinfo{booktitle}{\emph{EACL}}.
\newblock


\bibitem[\protect\citeauthoryear{Roy and Anand}{Roy and Anand}{2021}]%
        {saharoy2021question}
\bibfield{author}{\bibinfo{person}{Rishiraj~Saha Roy} {and}
  \bibinfo{person}{Avishek Anand}.} \bibinfo{year}{2021}\natexlab{}.
\newblock \bibinfo{booktitle}{\emph{{Question Answering for the Curated Web:
  Tasks and Methods in QA over Knowledge Bases and Text Collections}}}.
\newblock \bibinfo{publisher}{Morgan \& Claypool Publishers}.
\newblock


\bibitem[\protect\citeauthoryear{Saha, Pahuja, Khapra, Sankaranarayanan, and
  Chandar}{Saha et~al\mbox{.}}{2018}]%
        {saha2018complex}
\bibfield{author}{\bibinfo{person}{Amrita Saha}, \bibinfo{person}{Vardaan
  Pahuja}, \bibinfo{person}{Mitesh Khapra}, \bibinfo{person}{Karthik
  Sankaranarayanan}, {and} \bibinfo{person}{Sarath Chandar}.}
  \bibinfo{year}{2018}\natexlab{}.
\newblock \showarticletitle{Complex sequential question answering: Towards
  learning to converse over linked question answer pairs with a knowledge
  graph}. In \bibinfo{booktitle}{\emph{AAAI}}.
\newblock


\bibitem[\protect\citeauthoryear{Savenkov and Agichtein}{Savenkov and
  Agichtein}{2016}]%
        {savenkov2016knowledge}
\bibfield{author}{\bibinfo{person}{Denis Savenkov} {and}
  \bibinfo{person}{Eugene Agichtein}.} \bibinfo{year}{2016}\natexlab{}.
\newblock \showarticletitle{{When a knowledge base is not enough: Question
  answering over knowledge bases with external text data}}. In
  \bibinfo{booktitle}{\emph{SIGIR}}.
\newblock


\bibitem[\protect\citeauthoryear{Shang, Wang, Eric, Chen, Wang, Welch, Deng,
  Grewal, Wang, Liu, et~al\mbox{.}}{Shang et~al\mbox{.}}{2021}]%
        {shang2021entity}
\bibfield{author}{\bibinfo{person}{Mingyue Shang}, \bibinfo{person}{Tong Wang},
  \bibinfo{person}{Mihail Eric}, \bibinfo{person}{Jiangning Chen},
  \bibinfo{person}{Jiyang Wang}, \bibinfo{person}{Matthew Welch},
  \bibinfo{person}{Tiantong Deng}, \bibinfo{person}{Akshay Grewal},
  \bibinfo{person}{Han Wang}, \bibinfo{person}{Yue Liu}, {et~al\mbox{.}}}
  \bibinfo{year}{2021}\natexlab{}.
\newblock \showarticletitle{Entity Resolution in Open-domain Conversations}. In
  \bibinfo{booktitle}{\emph{NAACL}}.
\newblock


\bibitem[\protect\citeauthoryear{Shen, Geng, Tao, Guo, Tang, Duan, Long, and
  Jiang}{Shen et~al\mbox{.}}{2019}]%
        {shen2019multi}
\bibfield{author}{\bibinfo{person}{Tao Shen}, \bibinfo{person}{Xiubo Geng},
  \bibinfo{person}{QIN Tao}, \bibinfo{person}{Daya Guo}, \bibinfo{person}{Duyu
  Tang}, \bibinfo{person}{Nan Duan}, \bibinfo{person}{Guodong Long}, {and}
  \bibinfo{person}{Daxin Jiang}.} \bibinfo{year}{2019}\natexlab{}.
\newblock \showarticletitle{Multi-Task Learning for Conversational Question
  Answering over a Large-Scale Knowledge Base}. In
  \bibinfo{booktitle}{\emph{EMNLP-IJCNLP}}.
\newblock


\bibitem[\protect\citeauthoryear{Si, Zhao, and Boyd-Graber}{Si
  et~al\mbox{.}}{2021}]%
        {si2021whats}
\bibfield{author}{\bibinfo{person}{Chenglei Si}, \bibinfo{person}{Chen Zhao},
  {and} \bibinfo{person}{Jordan Boyd-Graber}.} \bibinfo{year}{2021}\natexlab{}.
\newblock \showarticletitle{{What's in a Name? Answer Equivalence For
  Open-Domain Question Answering}}. In \bibinfo{booktitle}{\emph{EMNLP}}.
\newblock


\bibitem[\protect\citeauthoryear{Suchanek, Kasneci, and Weikum}{Suchanek
  et~al\mbox{.}}{2007}]%
        {suchanek2007yago}
\bibfield{author}{\bibinfo{person}{Fabian Suchanek}, \bibinfo{person}{Gjergji
  Kasneci}, {and} \bibinfo{person}{Gerhard Weikum}.}
  \bibinfo{year}{2007}\natexlab{}.
\newblock \showarticletitle{{YAGO: A core of semantic knowledge}}. In
  \bibinfo{booktitle}{\emph{WWW}}.
\newblock


\bibitem[\protect\citeauthoryear{Sun, Bedrax-Weiss, and Cohen}{Sun
  et~al\mbox{.}}{2019}]%
        {sun2019pullnet}
\bibfield{author}{\bibinfo{person}{Haitian Sun}, \bibinfo{person}{Tania
  Bedrax-Weiss}, {and} \bibinfo{person}{William Cohen}.}
  \bibinfo{year}{2019}\natexlab{}.
\newblock \showarticletitle{PullNet: Open Domain Question Answering with
  Iterative Retrieval on Knowledge Bases and Text}. In
  \bibinfo{booktitle}{\emph{EMNLP-IJCNLP}}.
\newblock


\bibitem[\protect\citeauthoryear{Sun, Dhingra, Zaheer, Mazaitis, Salakhutdinov,
  and Cohen}{Sun et~al\mbox{.}}{2018}]%
        {sun2018open}
\bibfield{author}{\bibinfo{person}{Haitian Sun}, \bibinfo{person}{Bhuwan
  Dhingra}, \bibinfo{person}{Manzil Zaheer}, \bibinfo{person}{Kathryn
  Mazaitis}, \bibinfo{person}{Ruslan Salakhutdinov}, {and}
  \bibinfo{person}{William Cohen}.} \bibinfo{year}{2018}\natexlab{}.
\newblock \showarticletitle{Open Domain Question Answering Using Early Fusion
  of Knowledge Bases and Text}. In \bibinfo{booktitle}{\emph{EMNLP}}.
\newblock


\bibitem[\protect\citeauthoryear{Talmor, Yoran, Catav, Lahav, Wang, Asai,
  Ilharco, Hajishirzi, and Berant}{Talmor et~al\mbox{.}}{2021}]%
        {talmor2021multimodalqa}
\bibfield{author}{\bibinfo{person}{Alon Talmor}, \bibinfo{person}{Ori Yoran},
  \bibinfo{person}{Amnon Catav}, \bibinfo{person}{Dan Lahav},
  \bibinfo{person}{Yizhong Wang}, \bibinfo{person}{Akari Asai},
  \bibinfo{person}{Gabriel Ilharco}, \bibinfo{person}{Hannaneh Hajishirzi},
  {and} \bibinfo{person}{Jonathan Berant}.} \bibinfo{year}{2021}\natexlab{}.
\newblock \showarticletitle{MultiModalQA: Complex Question Answering over Text,
  Tables and Images}. In \bibinfo{booktitle}{\emph{ICLR}}.
\newblock


\bibitem[\protect\citeauthoryear{Thorne, Yazdani, Saeidi, Silvestri, Riedel,
  and Halevy}{Thorne et~al\mbox{.}}{2020}]%
        {thorne2020neural}
\bibfield{author}{\bibinfo{person}{James Thorne}, \bibinfo{person}{Majid
  Yazdani}, \bibinfo{person}{Marzieh Saeidi}, \bibinfo{person}{Fabrizio
  Silvestri}, \bibinfo{person}{Sebastian Riedel}, {and} \bibinfo{person}{Alon
  Halevy}.} \bibinfo{year}{2020}\natexlab{}.
\newblock \showarticletitle{Neural databases}.
\newblock \bibinfo{journal}{\emph{arXiv}} (\bibinfo{year}{2020}).
\newblock


\bibitem[\protect\citeauthoryear{Vakulenko, Longpre, Tu, and Anantha}{Vakulenko
  et~al\mbox{.}}{2021}]%
        {vakulenko2021question}
\bibfield{author}{\bibinfo{person}{Svitlana Vakulenko}, \bibinfo{person}{Shayne
  Longpre}, \bibinfo{person}{Zhucheng Tu}, {and} \bibinfo{person}{Raviteja
  Anantha}.} \bibinfo{year}{2021}\natexlab{}.
\newblock \showarticletitle{Question rewriting for conversational question
  answering}. In \bibinfo{booktitle}{\emph{WSDM}}.
\newblock


\bibitem[\protect\citeauthoryear{Voskarides, Li, Ren, Kanoulas, and
  de~Rijke}{Voskarides et~al\mbox{.}}{2020}]%
        {voskarides2020query}
\bibfield{author}{\bibinfo{person}{Nikos Voskarides}, \bibinfo{person}{Dan Li},
  \bibinfo{person}{Pengjie Ren}, \bibinfo{person}{Evangelos Kanoulas}, {and}
  \bibinfo{person}{Maarten de Rijke}.} \bibinfo{year}{2020}\natexlab{}.
\newblock \showarticletitle{Query resolution for conversational search with
  limited supervision}. In \bibinfo{booktitle}{\emph{SIGIR}}.
\newblock


\bibitem[\protect\citeauthoryear{Vrande{\v{c}}i{\'c} and
  Kr{\"o}tzsch}{Vrande{\v{c}}i{\'c} and Kr{\"o}tzsch}{2014}]%
        {vrandevcic2014wikidata}
\bibfield{author}{\bibinfo{person}{Denny Vrande{\v{c}}i{\'c}} {and}
  \bibinfo{person}{Markus Kr{\"o}tzsch}.} \bibinfo{year}{2014}\natexlab{}.
\newblock \showarticletitle{{Wikidata: A free collaborative knowledge base}}.
\newblock \bibinfo{journal}{\emph{CACM}} (\bibinfo{year}{2014}).
\newblock


\bibitem[\protect\citeauthoryear{Xiong, Yu, Chang, Guo, and Wang}{Xiong
  et~al\mbox{.}}{2019}]%
        {xiong2019improving}
\bibfield{author}{\bibinfo{person}{Wenhan Xiong}, \bibinfo{person}{Mo Yu},
  \bibinfo{person}{Shiyu Chang}, \bibinfo{person}{Xiaoxiao Guo}, {and}
  \bibinfo{person}{William~Yang Wang}.} \bibinfo{year}{2019}\natexlab{}.
\newblock \showarticletitle{Improving Question Answering over Incomplete KBs
  with Knowledge-Aware Reader}. In \bibinfo{booktitle}{\emph{ACL}}.
\newblock


\bibitem[\protect\citeauthoryear{Xu, Reddy, Feng, Huang, and Zhao}{Xu
  et~al\mbox{.}}{2016}]%
        {xu2016question}
\bibfield{author}{\bibinfo{person}{Kun Xu}, \bibinfo{person}{Siva Reddy},
  \bibinfo{person}{Yansong Feng}, \bibinfo{person}{Songfang Huang}, {and}
  \bibinfo{person}{Dongyan Zhao}.} \bibinfo{year}{2016}\natexlab{}.
\newblock \showarticletitle{Question Answering on Freebase via Relation
  Extraction and Textual Evidence}. In \bibinfo{booktitle}{\emph{ACL}}.
\newblock


\bibitem[\protect\citeauthoryear{Xu, Zhu, Geng, Yang, Lin, and Jiang}{Xu
  et~al\mbox{.}}{2020}]%
        {xu2020learning}
\bibfield{author}{\bibinfo{person}{Zihan Xu}, \bibinfo{person}{Jiangang Zhu},
  \bibinfo{person}{Ling Geng}, \bibinfo{person}{Yang Yang},
  \bibinfo{person}{Bojia Lin}, {and} \bibinfo{person}{Daxin Jiang}.}
  \bibinfo{year}{2020}\natexlab{}.
\newblock \showarticletitle{Learning to Generate Reformulation Actions for
  Scalable Conversational Query Understanding}. In
  \bibinfo{booktitle}{\emph{CIKM}}.
\newblock


\bibitem[\protect\citeauthoryear{Zhu, Lei, Huang, Wang, Zhang, Lv, Feng, and
  Chua}{Zhu et~al\mbox{.}}{2021}]%
        {zhu2021tat}
\bibfield{author}{\bibinfo{person}{Fengbin Zhu}, \bibinfo{person}{Wenqiang
  Lei}, \bibinfo{person}{Youcheng Huang}, \bibinfo{person}{Chao Wang},
  \bibinfo{person}{Shuo Zhang}, \bibinfo{person}{Jiancheng Lv},
  \bibinfo{person}{Fuli Feng}, {and} \bibinfo{person}{Tat-Seng Chua}.}
  \bibinfo{year}{2021}\natexlab{}.
\newblock \showarticletitle{TAT-QA: A Question Answering Benchmark on a Hybrid
  of Tabular and Textual Content in Finance}. In
  \bibinfo{booktitle}{\emph{ACL}}.
\newblock


\bibitem[\protect\citeauthoryear{Ziegler, Abujabal, Roy, and Weikum}{Ziegler
  et~al\mbox{.}}{2017}]%
        {ziegler2017efficiency}
\bibfield{author}{\bibinfo{person}{David Ziegler}, \bibinfo{person}{Abdalghani
  Abujabal}, \bibinfo{person}{Rishiraj~Saha Roy}, {and}
  \bibinfo{person}{Gerhard Weikum}.} \bibinfo{year}{2017}\natexlab{}.
\newblock \showarticletitle{Efficiency-aware Answering of Compositional
  Questions using Answer Type Prediction}. In
  \bibinfo{booktitle}{\emph{IJCNLP}}.
\newblock


\end{thebibliography}
